\documentclass[research]{fcs}
\usepackage{titlesec}
\usepackage{placeins}
\usepackage{microtype}
\usepackage{cite}
\usepackage{amsmath,amssymb,amsfonts}
\usepackage{algorithmic}
\usepackage{graphicx}
\usepackage{subfigure}
\usepackage{textcomp}
\usepackage{times}
\usepackage{epsfig}
\usepackage{bm}
\usepackage{color}
\usepackage{multirow}
\usepackage{xspace}
\usepackage{tabularx}
\usepackage{rotating}
\usepackage[export]{adjustbox}
\usepackage{gensymb}
\usepackage{ragged2e}
\usepackage{multicol}
\usepackage{stfloats}
\usepackage{xcolor}
\usepackage{colortbl}
\usepackage{float}

\newcommand{\redcite}[1]{\textcolor{black}{#1}}

\newcommand{\ie}{\textit{i.e.}\xspace}
\newcommand{\eg}{\textit{e.g.}\xspace}
\newcommand{\etal}{\textit{et al.}\xspace}

\newcommand{\resp}{\textit{resp.}\xspace}
\newcommand{\our}{{SIKD}\xspace}
\newcommand{\BL}{\textrm{BL}\xspace}

\newcommand{\revise}[1]{\textcolor{black}{#1}}
\usepackage{hyperref}
\hypersetup{
    colorlinks=true,
    linkcolor=blue,    % 使用自定义颜色
    filecolor=gray,
    urlcolor=red,     % 使用自定义颜色
    citecolor=blue,    % 使用自定义颜色
}

\title{Shape-intensity knowledge distillation for robust medical image segmentation}
\author[1,2,3,4]{Wenhui Dong}
\author[1,2,3,4]{Bo Du}
\author*[1,2,3,4]{Yongchao Xu}

\address[1]{National Engineering Research Center for Multimedia Software, School of Computer Science, Wuhan University, Wuhan 430072, China}
\address[2]{Institute of Artificial Intelligence, School of Computer Science, Wuhan University, Wuhan 430072, China}
\address[3]{Hubei Key Laboratory of Multimedia and Network Communication Engineering, Wuhan University, Wuhan 430072, China}
\address[4]{Medical Artificial Intelligence Research Institute of Renmin Hospital, Wuhan University, Wuhan 430072, China}
\fcssetup{
  received       = {month dd, yyyy},
  accepted       = {month dd, yyyy},
  corr-email     = {yongchao.xu@whu.edu.cn},
}
\begin{abstract}
Many medical image segmentation methods have achieved impressive results. Yet, most existing methods do not take into account the shape-intensity prior information. This may lead to implausible segmentation results, in particular for images of unseen datasets. In this paper, we propose a novel approach to incorporate joint shape-intensity prior information into the segmentation network. Specifically, we first train a segmentation network (regarded as the teacher network) on class-wise averaged training images to extract valuable shape-intensity information, which is then transferred to a student segmentation network with the same network architecture as the teacher via knowledge distillation. In this way, the student network regarded as the final segmentation model can effectively integrate the shape-intensity prior information, yielding more accurate segmentation results.
Despite its simplicity, experiments on five medical image segmentation tasks of different modalities demonstrate that the proposed Shape-Intensity Knowledge Distillation (\textit{\our}) consistently improves several baseline models (including recent MaxStyle and SAMed) under intra-dataset evaluation, and significantly improves the cross-dataset generalization ability. Code is available at \href{https://github.com/whdong-whu/SIKD}{https://github.com/whdong-whu/SIKD}.

\end{abstract}

\begin{document}

\section{Introduction}
\label{sec:introduction}
Medical image segmentation aims to predict the
semantic interpretation of each pixel, and is one of the crucial tasks
in clinical image analysis. Accurate and reliable automatic
segmentation is required to quickly provide clinicians with assistance
and advice to improve the efficiency of clinical workflows.

In the past decades, due to the emergence of deep learning~\cite{wang2022local, wang2022reslnet, xu2023weakly, zhang2023person, tan2023feature, guo2023cw, wu2023graph, ruan2023intellectual, ji2023teachers}, medical image segmentation~\cite{ronneberger2015u, cciccek20163d, zhou2018unet++, oktay2018attention, isensee2021nnu, shi2022local, billot2023synthseg} has witnessed substantial progress. The pioneer
U-Net~\cite{ronneberger2015u} presents skip-connection that
effectively fuses the shallow texture information and deep semantic
information, achieving very promising medical image segmentation
results in most cases. Though many variants of the
% U-Net~\cite{cciccek20163d, zhou2018unet++, oktay2018attention,
%   xiang2020bio, feng2020cpfnet, isensee2021nnu} have then been
U-Net~\cite{cciccek20163d, zhou2018unet++, oktay2018attention, isensee2021nnu} have then been proposed, U-Net still remains the popular \textit{de facto} network for medical images, achieving relatively good results compared to many of its alternatives~\cite{gut.2022.itmi}.

Inspired by the success of vision transformer (ViT) in image
recognition~\cite{dosovitskiy2021an} and semantic
segmentation~\cite{zheng2021rethinking, xie2021segformer}, some
transformer-based networks have been proposed for medical image
segmentation. Specifically,~\cite{chen2021transunet} propose TransUNet that replaces the
encoder in U-Net with a hybrid of ResNet and ViT, while keeping the
original decoder of U-Net for medical image segmentation. Based on the
popular Swin-transformer~\cite{liu2021swin},
Swin-Unet~\cite{cao2023swin}, a UNet-like pure transformer
architecture has been proposed. Both TransUNet and Swin-Unet achieve
very encouraging results in medical image segmentation.

\begin{figure}[t]
\centering

\includegraphics[width=0.98\linewidth]{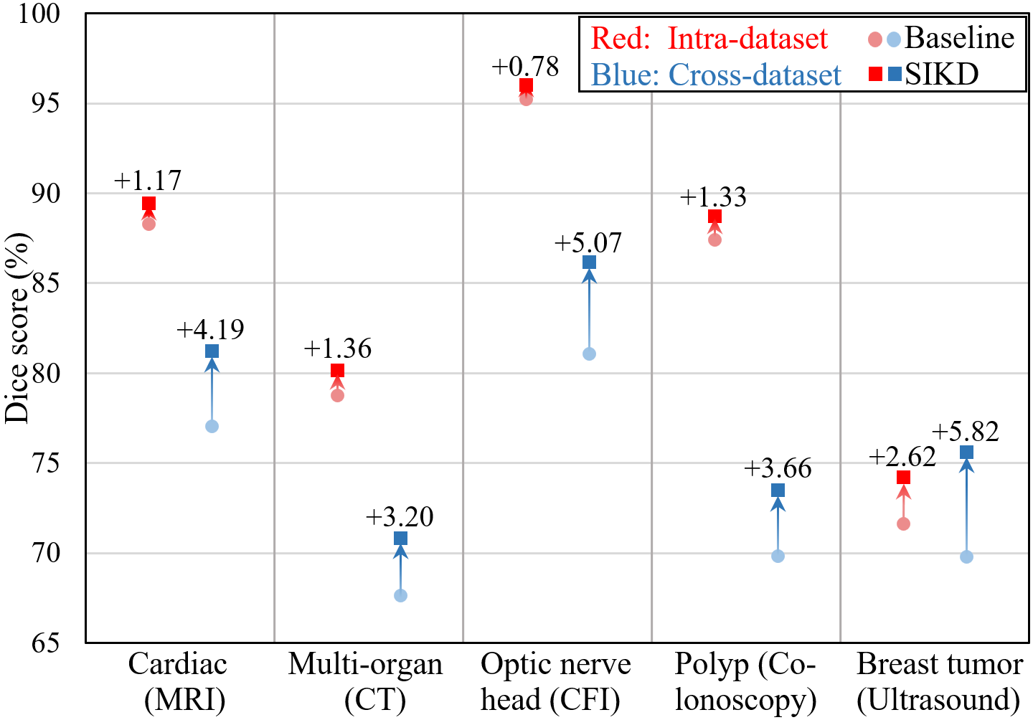}

\caption{
Comparison of averaged performance between the proposed \our and corresponding baseline models (including U-Net, SAUNet, PraNet, SANet, TransUnet, MaxStyle, SAMed, \revise{LM-Net and 2D D-LKA})
under intra-dataset and cross-dataset evaluation on five medical image
segmentation tasks of different modalities. The result is the average value across all baseline methods and the corresponding SIKD.
% The performance improvement is averaged across all tasks and baseline models, and is
% achieved with no extra runtime and memory cost during inference.
%% Illustration of the proposed \our on segmenting optic nerve head having elliptic shape. \our distillates the shape knowledge of the teacher model trained on the mean images from REFUGE dataset, and generalizes well to images from other datasets (e-f). Green outline: segmentation by the baseline model; Blue outline: segmentation by \our; Light blue area: ground-truth segmentation.
%Best viewed in the electronic version.
}

\label{fig:fig1}
\end{figure}

Most existing methods frame the medical image segmentation problem as
an individually pixel-wise classification task, and ignore that
medical objects usually have some specific shapes. This may lead to
anatomically implausible segmentation results, in particular for images of unseen datasets. Recently, some methods
% incorporating the shape prior information~\cite{sun2020saunet,
%   yan.2020.itmi, li.2022.mia.atrial, ning.2022.itmi,
%   oktay2017anatomically, ravishankar2017learning,
%   painchaud2020cardiac, larrazabal2020post, chen2021cooperative,
%   girum2021learning} to make the
incorporating the shape information~\cite{sun2020saunet,
  yan.2020.itmi, li.2022.mia.atrial, ning.2022.itmi,
  oktay2017anatomically, ravishankar2017learning,
 larrazabal2020post, chen2021cooperative} to make the
segmentation results more anatomically correct have been proposed. For
instance, some methods~\cite{sun2020saunet, yan.2020.itmi,
  li.2022.mia.atrial, ning.2022.itmi} learn extra shape-related
information and fuse shape feature and segmentation feature. Some other
methods~\cite{oktay2017anatomically, ravishankar2017learning,larrazabal2020post, chen2021cooperative} leverage the encoder-decoder network on both
predicted and ground-truth segmentation to align their latent non-linear representation, and refine the
predicted segmentation through auto-encoder.
% Adversarial learning is
% also used in~\cite{yi2019generative, JAFARI2022102536} to discriminate
% predicted segmentation and ground-truth segmentation, helping to get
% more plausible segmentation results.

% Organs or lesions in medical images often possess specific inherent anatomical structures. Hence, incorporating shape information can effectively improve the model’s performance and generalization capabilities. 
Incorporating shape information has been shown to improve
the performance of medical image segmentation. \revise{Most existing methods~\cite{sun2020saunet,
yan.2020.itmi, li.2022.mia.atrial, ning.2022.itmi,
oktay2017anatomically, ravishankar2017learning,
painchaud2020cardiac, larrazabal2020post, chen2021cooperative,
girum2021learning, 
zotti2018convolutional, tilborghs2022shape, mirikharaji2018star,
GUO2022102532, WEI2023102686}} force the network to learn and make use of the shape feature by predicting extra shape-related information or aligning non-linear representation of the segmentation result and ground truth. \revise{However, these methods often necessitate additional computations to fuse the learned shape features or a post-processing step to refine the segmentation. Furthermore, these methods fail to incorporate intensity information, which has been demonstrated to be valuable prior knowledge for medical image segmentation~\cite{yang20043d, wang2016shape}.}

In this paper, different from the existing methods, we propose incorporating joint shape-intensity information \revise{to enhance the model’s segmentation performance of the model, while simultaneously improving the model's generalization capabilities on unseen datasets.}
This is achieved by novelly leveraging knowledge distillation~\cite{hinton2015distilling}. 
% which is widely used for
% efficient image classification~\cite{hinton2015distilling, ba2014deep,
%   tian2020contrastive, wang.2022.pami, ye2022generalized} and
% segmentation~\cite{liu2020structured, wang2020intra, qin2021efficient,
%   shu2021channel, tian2022adaptive, yang2022cross}, cross-modal image
% analysis~\cite{gupta2016cross, hu2020knowledge, dou.2020.itmi,
%   li2020towards}, and efficient
% pre-training~\cite{he2022knowledge}. 
Specifically, we first train a segmentation network on the class-wise averaged training images without texture information. This segmentation network encodes the useful shape-intensity knowledge, and is regarded as the teacher network. We then train a student segmentation network with the same network architecture as the teacher network on the original training images. In addition to the classical segmentation loss, we also apply a distillation loss on the penultimate layer between the teacher and student network. In this way, the student network effectively learns shape-intensity information, leading to more plausible intra-dataset and cross-dataset segmentation results (see Fig.~\ref{fig:fig1}). The student network is considered as the final segmentation network, which does not require any extra computation cost during inference, making it efficient in practical usage. 

The main contribution of this paper is: 
1) \revise{We novelly propose to train a network  (\ie, teacher network) on the class-wise averaged training images. This simple design explicitly extracts shape-intensity prior information; }
% 1) Different from most existing deep-learning-based methods, we propose to incorporate joint shape-intensity prior information into the segmentation network. For that, we first extract shape-intensity prior information by training the segmentation network (regarded as the teacher model) on the class-wise averaged training images. %This simple design effectively extracts the shape-intensity feature for segmentation; 
2) We then leverage knowledge distillation to transfer
the shape-intensity knowledge learned by the teacher model to the student
network (\ie, final segmentation network)
, effectively incorporating
shape-intensity prior information for medical image segmentation; 
%To the best of our knowledge, this is for the first time that knowledge distillation is used to incorporate shape-intensity prior information for robust medical segmentation; 
3) Extensive experiments on five medical image segmentation tasks of different modalities, demonstrate that the proposed Shape-Intensity Knowledge Distillation (\our) consistently/significantly improves the baseline models and has a better generalization ability to images of unseen datasets.

The rest of this paper is organized as follows. In
Sec.~\ref{sec:related}, we review some related works on
medical image segmentation incorporating shape information and
knowledge distillation. We then detail the proposed method in
Sec.~\ref{sec:method}, followed by extensive experimental results in
Sec.~\ref{sec:experiment} and some discussions in Sec.~\ref{sec:discussion}. Finally, we conclude and give some perspectives in Sec.~\ref{sec:conclusion}.

\section{Related Work}
\label{sec:related}
Since the pioneer work of U-Net~\cite{ronneberger2015u}, there are
many U-Net-based works on medical image
segmentation~\cite{cciccek20163d, zhou2018unet++, oktay2018attention,
  xiang2020bio, feng2020cpfnet, isensee2021nnu, gut.2022.itmi}. A
detailed review of recent methods can
be found in~\cite{tajbakhsh2020embracing}. In this section, we mainly
focus on shortly reviewing some related works on deep-learning-based medical image
segmentation incorporating shape information and knowledge
distillation. More details about the shape-aware medical image
segmentation and knowledge distillation are referred to~\cite{
  xie2021survey},~\cite{wang2021knowledge} and~\cite{gou2021knowledge},
respectively.

\subsection{Shape-aware medical image segmentation}\label{subsec:shapeseg}

Many methods~\cite{sun2020saunet,
  yan.2020.itmi, li.2022.mia.atrial, ning.2022.itmi,
  oktay2017anatomically, ravishankar2017learning,
  painchaud2020cardiac, larrazabal2020post, chen2021cooperative,
  girum2021learning, 
  zotti2018convolutional, tilborghs2022shape, mirikharaji2018star,
  GUO2022102532, WEI2023102686} are proposed to incorporate the shape prior
information into segmentation network, and achieve more accurate segmentation results.

% Integrating prior knowledge into the segmentation network has proven
% useful to obtain more accurate segmentation
% results~\cite{xie2021survey}. Many methods~\cite{sun2020saunet,
%   yan.2020.itmi, li.2022.mia.atrial, ning.2022.itmi,
%   oktay2017anatomically, ravishankar2017learning,
%   painchaud2020cardiac, larrazabal2020post, chen2021cooperative,
%   girum2021learning, 
%   zotti2018convolutional, tilborghs2022shape, mirikharaji2018star,
%   GUO2022102532} are proposed to incorporate the shape prior
% information into segmentation network.

% In addition to the classical pixel-wise segmentation loss (\eg,
% cross-entropy loss or Dice loss), some methods~\cite{sun2020saunet,
%   yan.2020.itmi, li.2022.mia.atrial, ning.2022.itmi} also adopt other
% losses to learn extra shape-related target (usually based on the object
% boundary). 
% some methods~\cite{sun2020saunet, yan.2020.itmi, li.2022.mia.atrial, ning.2022.itmi} adopt other losses to learn extra shape-related targets (usually based on the object boundary). This helps to learn shape-related feature.
Some methods~\cite{sun2020saunet, yan.2020.itmi, li.2022.mia.atrial, ning.2022.itmi} rely on the adoption of additional losses to learn extra shape-related targets, which are typically based on the object boundary. 
For instance, SAUNet~\cite{sun2020saunet} adds a shape stream (supervised by ground-truth boundary) to the texture stream of U-Net with a decoder replaced by the spatial and channel-wise attention paths. A gated convolution layer is then used to fuse shape and texture information for segmentation. In~\cite{yan.2020.itmi}, the authors propose a loss based on the segment-level shape similarity that measures the curve similarity between each ground-truth boundary and corresponding predicted boundary segmentation, requiring no extra runtime during inference.
% Meanwhile, they have also proposed a more efficient framework to improve running speed.
AtrialJSQnet~\cite{li.2022.mia.atrial} predicts an additional distance map with respect to the boundary to incorporate spatial and shape information, without introducing additional inference time. In addition to classical segmentation and edge loss, SMU-Net~\cite{ning.2022.itmi} also adopts shape-aware loss characterizing the distance to the nearest object boundary and position-aware loss reflecting the distance to the object center.

Some methods~\cite{oktay2017anatomically, ravishankar2017learning,
  painchaud2020cardiac, larrazabal2020post, chen2021cooperative,
  girum2021learning} resort to apply an auto-reconstruction network
with encoder-decoder architecture on the predicted segmentation and
the ground-truth annotation. The encoder projects segmentation result
and annotation to latent non-linear representation, on which they
align the prediction and ground-truth segmentation. The decoder is
often used to refine the predicted
segmentation~\cite{ravishankar2017learning, painchaud2020cardiac,
  larrazabal2020post, chen2021cooperative, girum2021learning}, which
heavily relies on the ground-truth mask degradation strategy to train
the auto-reconstruction network. Specifically,~\cite{ravishankar2017learning} and
ACNN-Seg~\cite{oktay2017anatomically} share a similar idea of leveraging
such an auto-reconstruction pipeline. The latter does not rely on the
decoder to refine segmentation results, and thus needs no additional runtime during inference. Based on the similar
auto-reconstruction network,~\cite{painchaud2020cardiac} propose an anatomically-constrained
rejection sampling procedure to augment the latent representation, and
warp anatomically invalid segmentation toward the closest anatomically
plausible one. Post-DAE~\cite{larrazabal2020post} leverages denoising
autoencoders to post-process the segmentation result to anatomically
plausible segmentation. \redcite{Chen \etal}~\cite{chen2021cooperative} further
propose a hard example generation in the latent space of the
segmentation network to generate diverse training images and corrupted
segmentation results, reinforcing the performance of refined
segmentation. LFB-Net~\cite{girum2021learning} shares the same decoder
between the segmentation network and auto-reconstruction network, and
fuses the latent feature of both segmentation and auto-reconstruction
network.

Adversarial learning is also used to integrate shape
information~\cite{yi2019generative, JAFARI2022102536}, where the
segmentation network is regarded as the generator. The
core idea is to generate segmentation result that
confuses the discriminator in discriminating the ground-truth
segmentation and predicted segmentation. Jointly training the
discriminator and segmentation network helps to yield more plausible
segmentation results, without introducing any extra runtime during inference.

Some other methods~\cite{zotti2018convolutional, tilborghs2022shape,
  mirikharaji2018star, GUO2022102532} focus on fusing/learning prior
shape characteristics or segmenting objects of specific shapes (\eg,
star and circular shape). For instance, The method
in~\cite{zotti2018convolutional} fuses prior shape information about
the distribution of semantic class over the image domain
(statistically computed on the training set) into a segmentation
network. \redcite{Tilborghs~\etal}~\cite{tilborghs2022shape} directly learn
shape parameters of an underlying shape model statistically computed
on the set of training images, avoiding anatomically implausible
segmentation. \redcite{Mirikharaji and Hamarneh}~\cite{mirikharaji2018star}
introduce star shape regularized loss term to segment star shape skin
lesion, which does not require additional inference time. \redcite{Guo~\etal}~\cite{GUO2022102532} develop a globally optimal
label fusion (GOLF) algorithm that frames the predicted segmentation
and ``nesting''/circular shape priors into a normalize cut
framework, optimized by the proposed max-flow algorithm.

The existing methods that incorporate shape information achieve improved segmentation results. Most of them learn and fuse shape features guided by shape-related supervision, or rely on time-consuming auto-reconstruction to refine segmentation results. Some others devote to leveraging statistical shape models to take into account shape prior information. These methods often ignore the intensity prior information, which is also valuable for medical image segmentation~\cite{yang20043d, wang2016shape}. Differently, the proposed \our incorporates joint shape-intensity into deep-learning-based medical image segmentation. Besides, we also novelly resort to knowledge distillation to transfer the shape-intensity knowledge to the segmentation network. This further boosts the segmentation performance and generalization ability to unseen images.

\subsection{Knowledge distillation}
\label{subsec:kd}

Knowledge distillation (KD)~\cite{hinton2015distilling} generally
refers to transferring knowledge from a pre-trained teacher model to a
student model, to improve the performance of the student
network. Since the pioneer work~\cite{hinton2015distilling}, many
methods~\cite{ba2014deep, tian2020contrastive, wang.2022.pami,
  ye2022generalized,  GKD2023, 9975264} have been proposed for efficient image
classification. \redcite{Hinton~\etal}~\cite{hinton2015distilling} firstly
proposes that an efficient compact model could be obtained by
transferring the knowledge of the cumbersome model to the compact
model. \redcite{Ba and Caruana}~\cite{ba2014deep} also demonstrate through a series
of experiments that lightweight networks could learn complex functions
previously learned by deep networks. \redcite{Tian~\etal}~\cite{tian2020contrastive} suggest that previous KD methods ignore important
structural knowledge of the teacher network, and propose to leverage
contrastive learning to transfer structural knowledge.\redcite{Wang~\etal}~\cite{wang.2022.pami} discover that it is better to distillate
the knowledge lying in the penultimate layer of the teacher network
than mimicking the teacher's soft logits. Besides, the authors also
propose in~\cite{wang.2022.pami} to adopt locality-sensitive hashing
(LSH) to make the student focus more on mimicking the feature
direction than feature magnitude. REFILLED~\cite{ye2022generalized}
extends general knowledge distillation by applying both intra-image
relational KD and cross-image relational KD.

With the increasing efficiency requirement and hardware device limitation,
the idea of model compression via KD is gradually applied to other
vision tasks, such as semantic segmentation~\cite{liu2020structured,
  wang2020intra, qin2021efficient, shu2021channel, tian2022adaptive,
  yang2022cross}. For instance, in addition to classical pixel-wise
distillation loss, \redcite{Liu~\etal}~\cite{liu2020structured} propose
pair-wise distillation and holistic distillation (via discriminating
the segmentation of teacher and student network as real and fake,
respectively) loss, to make the student produce better structured
segmentation. \redcite{Wang~\etal}~\cite{wang2020intra} propose to compute
the class-wise feature prototype (class-wise averaged feature) and
then leverage knowledge distillation to transfer the intra-class
feature variation of the cumbersome teacher model to the compact
student model. \redcite{Qin~\etal}~\cite{qin2021efficient} propose to distillate
regional affinity between class-wise averaged feature prototype for
efficient medical image segmentation. \redcite{Shu~\etal}~\cite{shu2021channel}
improve KD-based segmentation by normalizing the activation map of
each feature channel to a soft probability map before minimizing the
Kullback–Leibler (KL) divergence between the teacher and student
model. This makes the distillation process pay more attention to the
most salient feature, which is valuable for segmentation. The adaptive
perspective distillation~\cite{tian2022adaptive} distills inter- and
intra- distribution of cosine similarity between the adapted feature
and adapted feature prototype. CIRKD~\cite{yang2022cross} distills the
cross-image relational knowledge to transfer global pixel
relationships between images for segmentation.

\begin{figure*}[tp]
\centering
\includegraphics[width=0.9\linewidth]{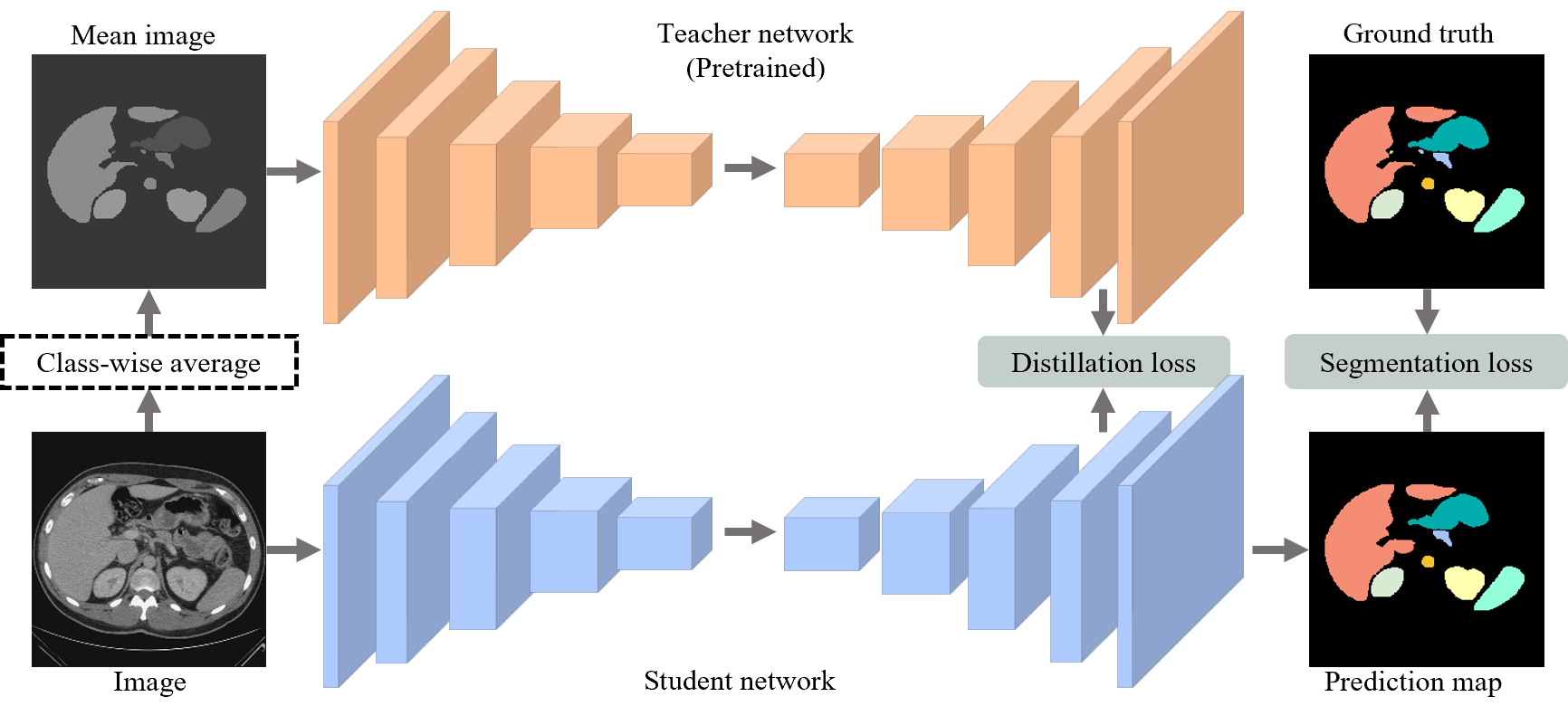}
%\vspace{\fixedvskip}
%\vspace{-6mm}
\caption{The pipeline of the proposed method. The teacher and student
  model have the same network architecture, trained on class-wise averaged
  training images and original images with segmentation loss, respectively. For the student model, we also 
 apply the distillation loss on the penultimate layer between the teacher and student
  model to transfer the shape-intensity knowledge.}
\label{fig:pipeline}
\end{figure*}

KD is also widely used for cross-modal image
analysis~\cite{gupta2016cross, hu2020knowledge, dou.2020.itmi,
  li2020towards}. For example, \redcite{Gupta~\etal}~\cite{gupta2016cross} propose feature mimicking between teacher
and student model (with the same network architecture) for
transferring learned representation from a large labeled modality to a
new unlabeled modality, enabling to learn rich representation from
unlabeled modalities. In~\cite{hu2020knowledge}, the authors train the
teacher model on concatenated multi-modal images, and leverage KD to
transfer multi-modal knowledge to mono-modal segmentation network. \redcite{Dou~\etal}~\cite{dou.2020.itmi} minimize KL-divergence of the confusion
matrix of class-wise averaged soft logits between different
modalities to achieve unpaired multi-modal segmentation via
KD. \redcite{Li~\etal}~\cite{li2020towards} propose an online mutual
knowledge distillation for cross-modality medical image segmentation,
where the segmentor on each modality explores the knowledge of another
modality via mutual KD. 
% He \etal~\cite{he2022knowledge} propose a
% non-parametric feature distillation method that allows the student
% network to be better adapted to downstream tasks using pre-trained
% models.

Most KD methods transfer knowledge between different networks or
modalities. There are also some self-distillation
methods~\cite{hou.2019.iccv, zhang.2022.pami} that transfer knowledge
of deep layers to shallow layers of the same network applied on the
same modality. Specifically, SAD~\cite{hou.2019.iccv} performs
top-down and layer-wise activation-based attention distillation within
the network itself, achieving effective lightweight lane
detection. \redcite{Zhang~\etal}~\cite{zhang.2022.pami} apply several
shallow classifiers at different shallow layers of the same network,
and make shallow classifiers mimic the deep classifier.
% , leading to an effective and compact neural network.

The existing KD methods are mainly developed for efficient image
classification and semantic segmentation, cross-modal image analysis. Differently, we
novelly leverage KD for incorporating shape-intensity knowledge for medical
image segmentation. Both the teacher and student model have
the same network architecture, trained on class-wise averaged images
without texture information and original training images,
respectively. The most related works~\cite{dou.2020.itmi,li2020towards} adopt KD to transfer common shape knowledge between different medical image modalities. The proposed \our focuses on extracting joint shape-intensity knowledge from single modality, and incorporating such prior information to achieve accurate and robust medical image segmentation of the underlying modality. %It is also noteworthy that the mode of using teacher and student network is also involved in mean-teacher framework~\cite{xu2021noisy} for addressing noisy labels, where the teacher network is updated from the student using Exponential Moving Average (EMA). Therefore, though such mean-teacher framework involve a teacher and student model concept, they are substantially different. We have now made our contributions clearer in the current version of the paper.

% To the best of our knowledge, we are the first to adopt
% knowledge distillation for incorporating shape knowledge in the
% segmentation network, leading to more accurate medical image
% segmentation.

\section{Proposed Method}
\label{sec:method}

\subsection{Motivation}
\label{subsec:overview}
Medical image segmentation has recently witnessed great progress
thanks to the development of deep learning. Most methods frame the
problem as individually pixel-wise classification, and adopt
U-Net~\cite{ronneberger2015u} or its various alternatives to learn
feature representation for pixel-wise classification. Such classical
learning-based segmentation scheme usually ignores the fact that the
object of interest in medical images generally has a specific shape, and 
the intensity prior information is also useful for medical image segmentation.
Indeed, it is usually easier to segment objects of
relatively smooth appearance than textured objects from an image (see
Fig.~\ref{fig:pipeline} for an example)). The smoothed object mainly
contains specific shape and homogeneous intensity information other than texture
information.

To incorporate the shape-intensity information into the medical image segmentation, we propose to first extract the shape-intensity knowledge from class-wise averaged training images, then transfer such prior knowledge using knowledge distillation~\cite{hinton2015distilling}. Specifically, for each training image $f$, we first compute the class-wise averaged image
$f_m$, which has only the shape-intensity information. We then feed $f_m$ to the teacher segmentation network to extract the shape-intensity information. Knowledge distillation is then adopted to transfer the prior knowledge to the student segmentation network, which has the same network architecture as the teacher network but with the original training images as inputs. In this way, the student model effectively acquires shape-intensity knowledge and is regarded as the final segmentation model. The proposed Shape-Intensity Knowledge Distillation (\our) pipeline is depicted in Fig.~\ref{fig:pipeline}.

\begin{figure}[tp]
\centering
\subfigure[On original images]{
% minipage中放置4张图，通过vspace进行换行调整，但是都在subfigure下面，共享一个子标题
\begin{minipage}[b]{0.48\linewidth}
\includegraphics[width=1\linewidth]{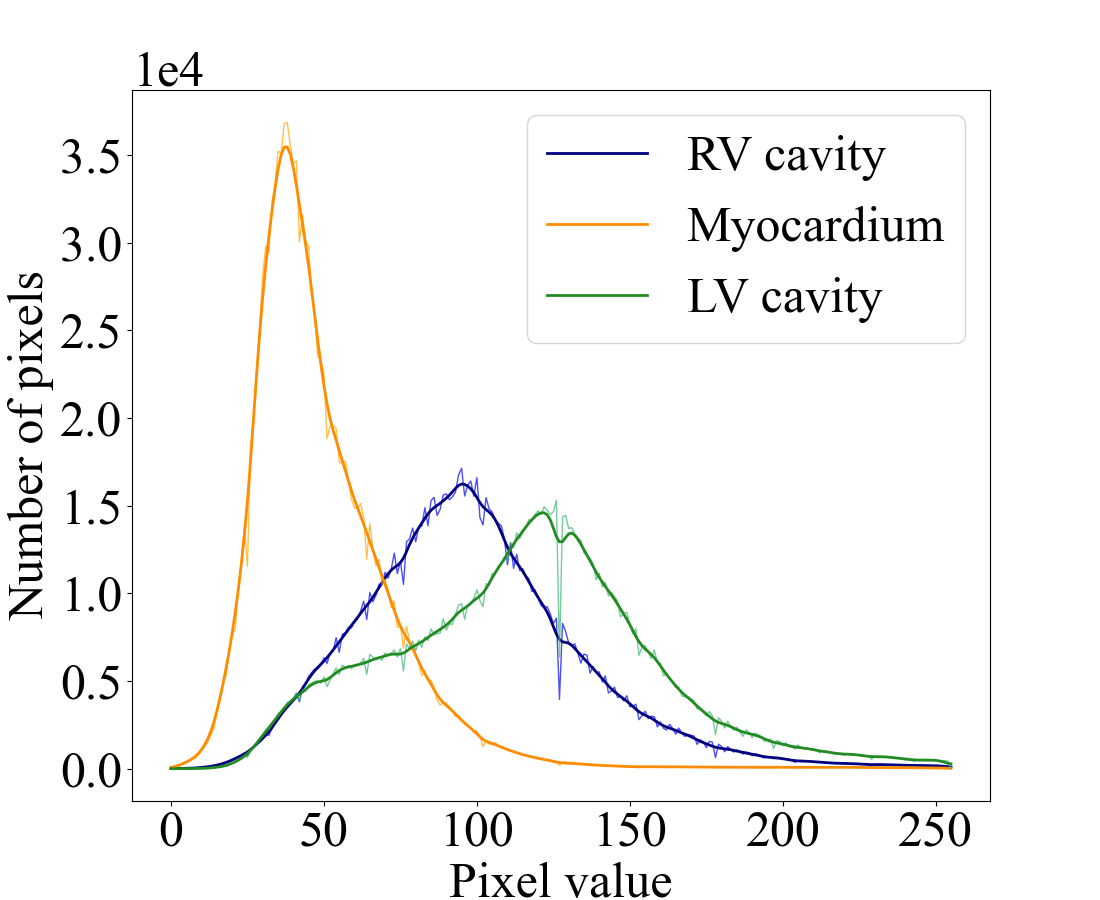}\vspace{3pt}
\end{minipage}}\hspace{-1.3mm}
\subfigure[On class-wise averaged images]{
\begin{minipage}[b]{0.48\linewidth}
\includegraphics[width=1\linewidth]{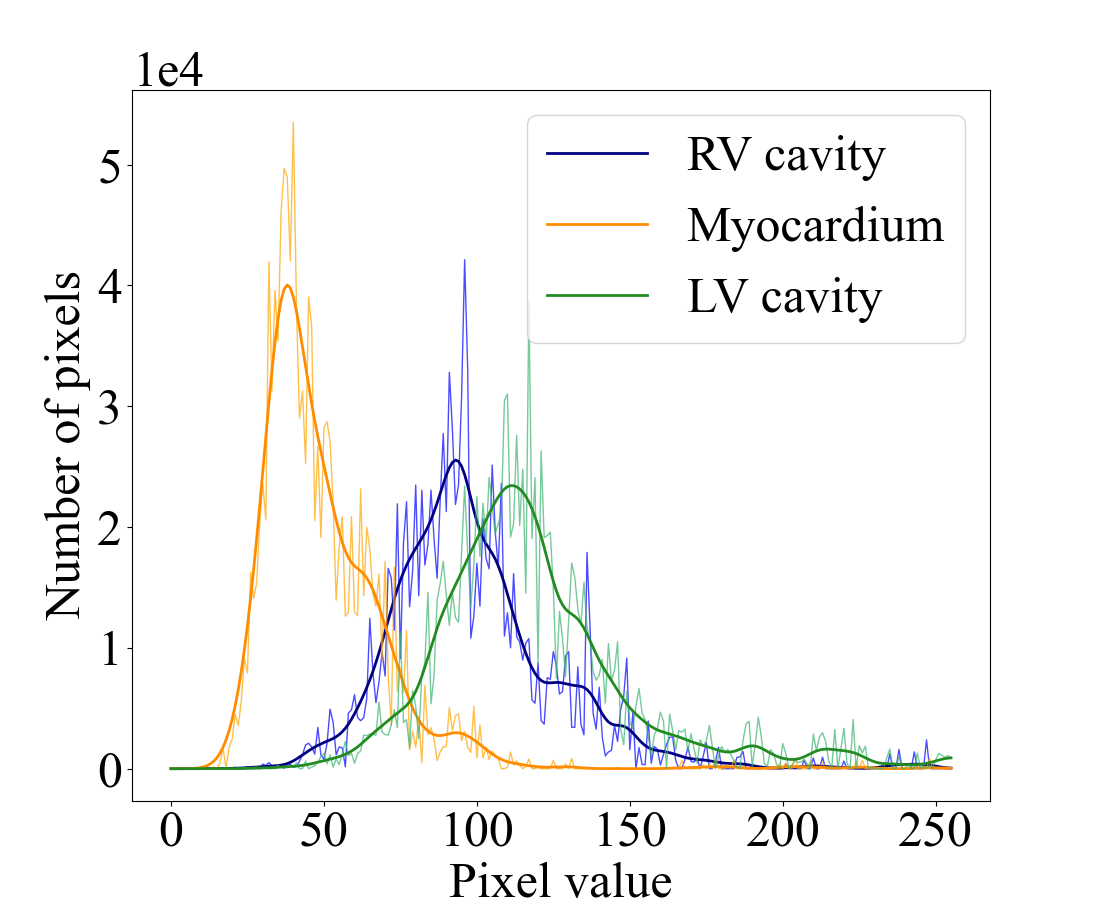}\vspace{3pt}
\end{minipage}}\hspace{-1.3mm}
	\caption{Distribution of pixel intensity within each class (\eg, RV cavity, Myocardium, and LV cavity) of the original images (a) and class-wise averaged images (b) on the ACDC training dataset~\cite{bernard2018deep}.}
\label{fig:ACDC_intensity}
\end{figure}

\subsection{Shape-intensity knowledge encoding}
\label{subsec:Encodeshape}
The object of interest in medical images is usually intensity inhomogeneous, and contains some texture information. On the other hand, objects in medical images often have a specific shape. Therefore, medical image segmentation can benefit from the
shape and intensity information. To encode such shape-intensity information, we propose to apply a segmentation network on the class-wise averaged training images.
More precisely, for a given image $f$, we calculate the class-wise averaged image $f_m$ by taking the mean value of the pixel values within each class. Formally, let $X_k$ denote the set of pixels belonging to the $k$-th class. For each pixel $x \in X_k$, the class-wise averaged image $f_m(x)$ is given by:
\begin{equation}
  \label{eq:meanimage}
  \begin{split}
    f_m(x) \, = \, \sum_{x \in X_k} f(x)/{\left|X_k\right|},
  \end{split}
\end{equation}
where $|\cdot|$ denotes the cardinality. It is noteworthy that when the input image is in RGB format, we calculate the class-wise averaged value separately for each channel.
Such class-wise averaged image $f_m$ has two distinct characteristics: 1) It does not contain any texture information (see Fig.~\ref{fig:pipeline}); 2) Its pixel value distribution is consistent with that of the original image (see Fig.~\ref{fig:ACDC_intensity}). Thanks to these two characteristics, training a segmentation network on this class-wise averaged image enables the network to learn shape-intensity information. We regard such a trained segmentation network as the teacher network.
% the learned feature is not significantly different from the same segmentation network trained on the original images.

We train the teacher network on the class-wise averaged training
images with classical segmentation loss
$\mathcal{L}_{seg}$. Specifically, for the U-Net based \our, we adopt
cross-entropy loss. For the other alternatives of \our, we adopt the
same cross-entropy loss and Dice loss with the corresponding baseline
models as the segmentation loss.

\subsection{Shape-intensity knowledge distillation}
\label{subsec:Shapedistillation}
Knowledge distillation is usually used to transfer knowledge from a cumbersome convolutional neural network (teacher) to a compact convolutional neural network (student) by aligning the representation of some layers of two networks. In this way, the compact student network is equipped with the powerful feature representation of the teacher network, which helps to improve the performance of the student network. Unlike these classical knowledge distillation frameworks, we propose to leverage knowledge distillation to transfer the shape-intensity information extracted by the teacher network described in Sec.~\ref{subsec:Encodeshape}. Specifically, the student network regarded as the final segmentation network has the same network architecture as the teacher network. We train the student network on the original training images.
%% Inspired by the basic ideas of knowledge distillation,
%% we propose to have the teacher network encode the shape representation
%% of the input image by calculating the mean value for each category of the
%% input and let the student network to learn this knowledge with the goal
%% that the student network can learn a better representation.
%% However, unlike most knowledge distillation methods, which
%% choose a cumbersome segmentation network as the teacher network
%% and a compact segmentation network as the student network.
In addition to the classical segmentation loss function, we also adopt a distillation loss to align the feature of the penultimate layer of both the teacher and student network. Let $f^t$ and $f^s$ denote the feature of the penultimate layer (before the $1 \times 1$ segmentation layer) of the teacher and student network, respectively. The adopted knowledge distillation loss $\mathcal{L}_{kd}$ is given by:
\begin{equation}
\label{eq:losskd}
\begin{split}
\mathcal{L}_{kd} \, &= \, MSE(f^t, \;f^s),
\end{split}
\end{equation}
where MSE($\cdot $) is the Mean-Squared Error loss function.  
% Since
% both the teacher and student network have the same network
% architecture, and are trained on similar images, 
This distillation loss facilitates the shape-intensity knowledge transfer from the teacher segmentation network to the student segmentation network.

For the student network trained on the original training images, the whole training objective function $\mathcal{L}$ consists of the same segmentation loss as the teacher segmentation network and the knowledge distillation loss defined in Eq.~\eqref{eq:losskd}. Formally, $\mathcal{L}$ is given by:
\begin{equation}
\label{eq:losstotal}
\begin{split}
  \mathcal{L} \, &= \, \mathcal{L}_{seg} \ + \ \alpha \times \mathcal{L}_{kd},
\end{split}
\end{equation}
where $\alpha$ is a hyper-parameter (set to 2 in all experiments) to
control the contribution of segmentation loss and distillation loss
term. It is noteworthy that the teacher segmentation network and the
knowledge distillation process are only involved in the training
phase. Therefore, during the inference phase, the segmentation network
does not require any extra runtime and memory cost.

%Given a well pre-trained teacher network using mean images, then the parameters of the student network are optimized by the end-to-end method according to the loss function Eq. \ref{eq:losstotal}. The teacher network and distillation module are deprecated in the inference stage.

\section{Experiments}
\label{sec:experiment}

%% We evaluate the proposed method on multiple benchmarks:
%% DOTA~\cite{xia2018dota} and HRSC2016~\cite{liu2016ship} for object
%% detection in aerial images, MSRA-TD500~\cite{yao2012detecting} and
%% RCTW-17~\cite{shi2017icdar2017} for long and oriented scene text
%% detection, and MW-18Mar~\cite{mw18mar} for pedestrian detection in
%% fisheye images. The ablation study is conducted on
%% DOTA~\cite{xia2018dota}, which is a challenging dataset for
%% multi-oriented object detection.

%This section introduces the datasets and quantitative metrics used for experiments first. Then describe the implementation details of the proposed method and ablation studies using different tasks.

We conduct \revise{intra-} and cross-dataset experiments on five different segmentation tasks of different medical imaging modalities to demonstrate the effectiveness of the proposed \our. The test set of intra-dataset serves as a kind of validation set to select the model. We mainly focus on cross-dataset segmentation performance using the selected model.

\subsection{Datasets and evaluation protocols}
\label{subsec:datasets}

% We conduct experiments on five different medical image segmentation
% tasks on different medical imaging modalities to demonstrate the
% effectiveness of the proposed \our. The details of the corresponding
% segmentation tasks and involved datasets are given in the following.

\smallskip
\noindent
\textbf{Cardiac segmentation in MRI images:}
Automated Cardiac Diagnosis Challenge
(ACDC)~\cite{bernard2018deep} releases
100 annotated MRI volumes obtained from 100 different patients.
% by two MRI scanners (1.5T and 3.0T)
% from 100 different patients. 
% The goal is to segment left ventricle
% (LV), right ventricle (RV), and myocardium (MYO). 
We randomly divide the dataset into 7:1:2 for training, validation, and test,
respectively. 
% For all methods, we report the results for the test set
% using the best model of different training epochs on the validation
% set.
We also evaluate the corresponding models on the 
training set of M\&Ms dataset~\cite{campello2021multi} to
evaluate the generalization ability of different methods. M\&Ms
releases 150 annotated training images from two different MRI vendors.
%(75 each).

\smallskip
\noindent
\textbf{Multi-organ segmentation in CT images:}
% We then conduct experiments on 
The Synapse multi-organ segmentation
dataset~\cite{landman2015miccai}
contains 30 abdominal CT scans, where 18 (\resp 12) cases are used for
training (\resp testing).
% involved in the MICCAI 2015 Multi-Atlas Abdomen Labeling
% Challenge. This dataset contains 3779 axial contrast-enhanced images
% from 30 abdominal CT scans, where 18 (\resp 12) cases are used for
% training (\resp testing). Each image slice is of size $512
% \times 512$.
%% The spatial resolution of each scan is $([0.54
%%   \sim 0.54] \times [0.98 \sim 0.98] \times [2.5 \sim 5.0])$
%% $\textrm{mm}^3$.
The goal is to segment 8 abdominal organs.
% : aorta, gallbladder, left
% kidney, right kidney, liver, pancreas, spleen, and stomach. 
% Following TransUnet~\cite{chen2021transunet}, we report the result of the model
% given by the last epoch of TransUnet and \our based on TransUnet. For
% the other methods, we report results of the best model on the test
% set.
We also conduct evaluations on the AbdomenCT-1K
dataset~\cite{ma2021abdomenct} containing more 1000 CT
scans from 12 medical centers to benchmark the generalization ability of different methods. 
% for liver, kidney, spleen, and pancreas
% segmentation.

\smallskip
\noindent
\textbf{Polyp segmentation in colonoscopy images:} Following the
experimental setup in~\cite{fan2020pranet}, we conduct experiments on
five public datasets for colorectal polyp segmentation, including
Kvasir~\cite{jha2020kvasir}, CVC-ClinicDB~\cite{bernal2015wm},
CVC-ColonDB~\cite{tajbakhsh2015automated},
ETIS~\cite{silva2014toward}, and
Endoscene~\cite{vazquez2017benchmark}.
% The Kvasir and CVC-ClinicDB dataset for training and testing, respectively.
We randomly split Kvasir and CVC-ClinicDB
dataset into 4:1 for training and testing, respectively. 
The CVC-ColonDB, ETIS~\cite{silva2014toward}, and Endoscene~\cite{vazquez2017benchmark} are used for assessing the generalization ability of different methods.
% We randomly split Kvasir and
% CVC-ClinicDB dataset into 4:1 for training and testing,
% respectively. We also perform evaluations on the other three datasets
% to assess the generalization ability of different methods. 
Note that the Endoscene dataset is a combination of CVC-ClinicDB and CVC-300. We
only evaluate the model on the test set of CVC-300 denoted as
CVC-T. 
% For all methods, we report results of the best model of
% different tr

\smallskip
\noindent
\textbf{Optic nerve head segmentation in color fundus images:}
% We then conduct experiments on segmenting optic nerve head (ONH) in color
% fundus images (CFI). The ONH usually has an elliptic shape. We train
% the segmentation model on 
The REFUGE challenge dataset~\cite{orlando2020refuge}
contains 400 color fundus images (CFI), randomly divided into 4:1 for training and
testing, respectively. 
To further demonstrate the generalization
ability of the proposed \our, we also evaluate different models
(trained on REFUGE dataset) on the public
Drishti-GS~\cite{sivaswamy2015comprehensive} dataset consisting of 101
images and RIM-ONE-r3~\cite{fumero2011rim} dataset containing 159
images. 
% For all methods, we report the result of the best model of
% different training epochs on the test set of REFUGE.
%datasets as test sets to validate the generalization of our method.

\smallskip
\noindent
\textbf{Breast tumor segmentation in ultrasound images:}
The BUSI dataset~\cite{al2020dataset} consists of a total of 780 images, including 487 benign images, 210 malignant images, and 133 normal images.
This dataset is randomly split into 4:1 for training and
testing. 
% For all methods, we report results of the best model on the
% test set. 
 We also conduct
the cross-dataset evaluation on the dataset used in~\cite{zhuang2019rdau} and the dataset B~\cite{yap.datasetB} to benchmark the generalization ability of different methods. The 
dataset in~\cite{zhuang2019rdau} is composed of 42 breast ultrasound
(BUS) images.
% provided by the Imaging Department of the First
% Affiliated Hospital of Shantou University, Shantou (China). 
The dataset B~\cite{yap.datasetB} consists of 163 images corresponding to 110 benign and 53 malignant lesion images.

\medskip
\noindent
\textbf{Evaluation protocols} We adopt three metrics: Dice score
(Dice), Intersection over Union (IoU), and Hausdorff Distance (HD),
which are widely used metrics in medical image segmentation to
evaluate the proposed \our. 
\begin{table*}[tp]
	\centering
	  % \normalsize
	\caption{Quantitative comparison between the baseline models and the proposed \our on ACDC dataset~\cite{bernard2018deep} and M\&Ms dataset~\cite{campello2021multi} for cardiac segmentation in MRI image.
 % $\dag$ represents the reproduced results based on the open-sourced implementation.
          %We also report results on the corrupted ACDC dataset~\cite{chen2021cooperative}.
          %We mainly rely on Dice score and HD (mm) to benchmark different methods. 
          % Note that all models are trained on the 70 training subjects of ACDC dataset.
		$\uparrow$ (\resp $\downarrow$) indicates the higher (\resp lower) the better.
	}
	\label{tab:ACDC_2}
	% \vspace{\fixedvskip}
 
	\setlength{\tabcolsep}{1.6mm}{
         \footnotesize%\small
 
		\begin{tabular}{lcccccccccccccccc}
			\toprule
			\multirow{5}{*}{Method} &
			\multicolumn{8}{c}{Intra-dataset: ACDC} &
			\multicolumn{8}{c}{Cross-dataset: M\&Ms} \\
			\cmidrule(r){2-9}
                \cmidrule(l){10-17}
			&
			\multicolumn{2}{c}{Average} &
			\multicolumn{2}{c}{LV} &
			\multicolumn{2}{c}{Myo} &
			\multicolumn{2}{c}{RV} &
			\multicolumn{2}{c}{Average} &
			\multicolumn{2}{c}{LV} &
			\multicolumn{2}{c}{Myo} &
			\multicolumn{2}{c}{RV} \\
			\cmidrule(r){2-3}
                \cmidrule(lr){4-5}
                \cmidrule(lr){6-7}
                \cmidrule(lr){8-9}
                \cmidrule(lr){10-11}
                \cmidrule(lr){12-13}
                \cmidrule(lr){14-15}
                \cmidrule(l){16-17}

			& DICE$\uparrow$ &HD$\downarrow$  & DICE &HD & DICE &HD & DICE &HD & DICE &HD & DICE &HD & DICE &HD & DICE &HD\\
			\midrule
			U-Net~\cite{ronneberger2015u} &0.884 &1.30 &0.905 &1.47 &0.852 &1.24 &0.895 &1.20 &0.686 &4.08 &0.733 &3.26 &0.641 &4.88 &0.684 &4.09 \\
U-Net w/ SDM~\cite{xue2020shape} &0.867 &1.54 &0.914 &1.25 &0.837 &2.09 &0.850 &1.29 &0.703 &4.54 &0.753 &4.65 &0.658 &6.74 &0.697 &3.22 \\
U-Net + \our &0.894 &1.21 &0.925 &1.29 &0.861 &1.10 &0.895 &1.23 &0.785 &2.95 &0.831 & 2.82 &0.741 &3.57 &0.783 &2.48\\
\midrule
Baseline (\BL) in~\cite{sun2020saunet} &0.870 &2.49 &0.913 &2.33 &0.837 &3.62 &0.860 &1.52 &0.729 &7.77 &0.796 &9.73 &0.662 &7.53 &0.728 &6.05 \\
SAUNet~\cite{sun2020saunet} &0.880 &2.05 &0.910 &1.88 &0.845 &2.31 &0.883 &1.96 &0.739 &4.57 &0.806 &5.93 &0.672 &4.21 &0.738 &3.56\\
\BL + \our &0.892 &1.58 &0.924 &1.36 &0.856 &1.93 &\textbf{0.897} &1.45 &0.766 &3.73 &0.816 &4.39 &0.733 &3.67 &0.750 &3.13\\
\midrule
MaxStyle~\cite{chen2022maxstyle} &0.876 &2.23 &0.914 &2.56 &0.843 &2.28 &0.872 &1.85 &0.818 &3.84 &0.826 &3.17 &0.780 &5.62 &0.847 &2.74 \\
MaxStyle + \our &0.888 &1.95 &0.923 &2.05 &0.857 &1.87 &0.883 &1.92 &0.827 &2.85 &0.833 &2.81 &0.792 &3.13 &\textbf{0.857} &2.62\\
\midrule
SAMed~\cite{zhang2023customized} &0.890 &1.15 &0.924 &1.11 &0.853 &1.12 &0.894 &1.22 &0.826 &2.38 &0.862 &2.27 &0.783 &2.39 &0.833 &2.46\\
SAMed + \our &0.893 &\textbf{1.10} &0.929 &\textbf{1.04} &0.855 &1.11 &0.895 &1.16 &0.842 &2.33 &0.873 &2.29 &0.799 &1.89 &0.855 &2.81\\
\midrule
\revise{LM-Net~\cite{LU2024107717}}&0.881 &1.53&0.922 &1.21 &0.852 &1.44  &0.867 &1.93 &0.742 &3.17 &0.828 &3.11 &0.713 &3.97 &0.684 &\textbf{2.43} \\
\revise{LM-Net + \our} &0.900 &1.33 &0.933 &1.28 &0.873 &1.41 &0.894 &\textbf{1.11} &0.802 &3.12 &0.866 &2.85 &0.774 &2.70 &0.767 &3.80 \\
\midrule
\revise{2D D-LKA Net~\cite{azad2024beyond}}  &0.898 &2.06 &0.929 &1.69 &0.875 &\textbf{1.09} &0.889 &3.39 &0.824 &2.77 &0.881 &2.03 &0.778 &1.84 &0.813 &4.43 \\
\revise{2D D-LKA + \our}  &\textbf{0.902} &1.91 &\textbf{0.933} &1.51 &\textbf{0.875} &1.23 &0.896 &2.98 &\textbf{0.854} &\textbf{2.21} &\textbf{0.905} &\textbf{1.75} &\textbf{0.822} &\textbf{1.36} &0.835 &3.51 \\
			\bottomrule
		\end{tabular}
		% \vspace{\fixedvskip}
	}
	%\vspace{2mm}
\end{table*}

\begin{figure}[tp]
\centering
% \begin{minipage}{1\linewidth}
% \centering
% \includegraphics[width=0.9\linewidth]{figure/ACDC/ACDC_color_bar_v2.png}
% \end{minipage}\vspace{-3pt}

\subfigure[Image]{
% minipage中放置4张图，通过vspace进行换行调整，但是都在subfigure下面，共享一个子标题
\begin{minipage}[b]{0.235\linewidth}
\includegraphics[width=1\linewidth]{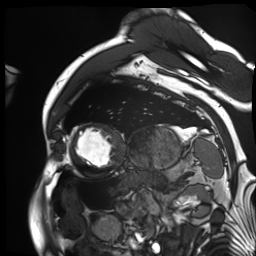}\vspace{3pt}
\includegraphics[width=1\linewidth]{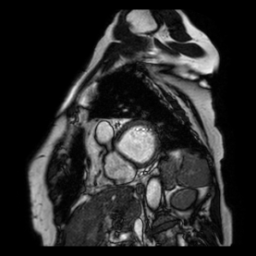}\vspace{3pt}
\includegraphics[width=1\linewidth]{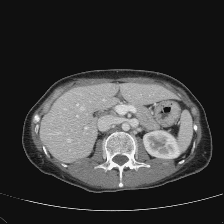}\vspace{3pt}
\includegraphics[width=1\linewidth]{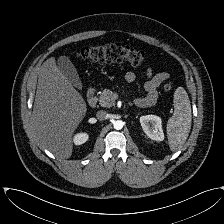}
\end{minipage}}\hspace{-1.3mm}
\subfigure[Baseline]{
\begin{minipage}[b]{0.235\linewidth}
\includegraphics[width=1\linewidth]{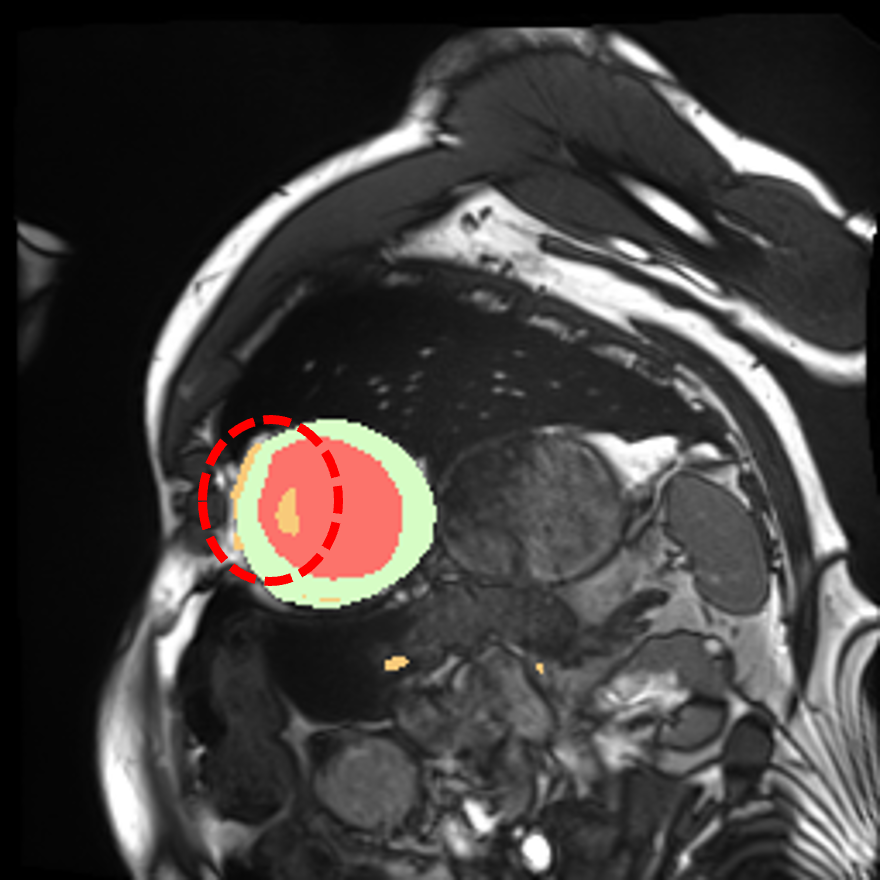}\vspace{3pt}
\includegraphics[width=1\linewidth]{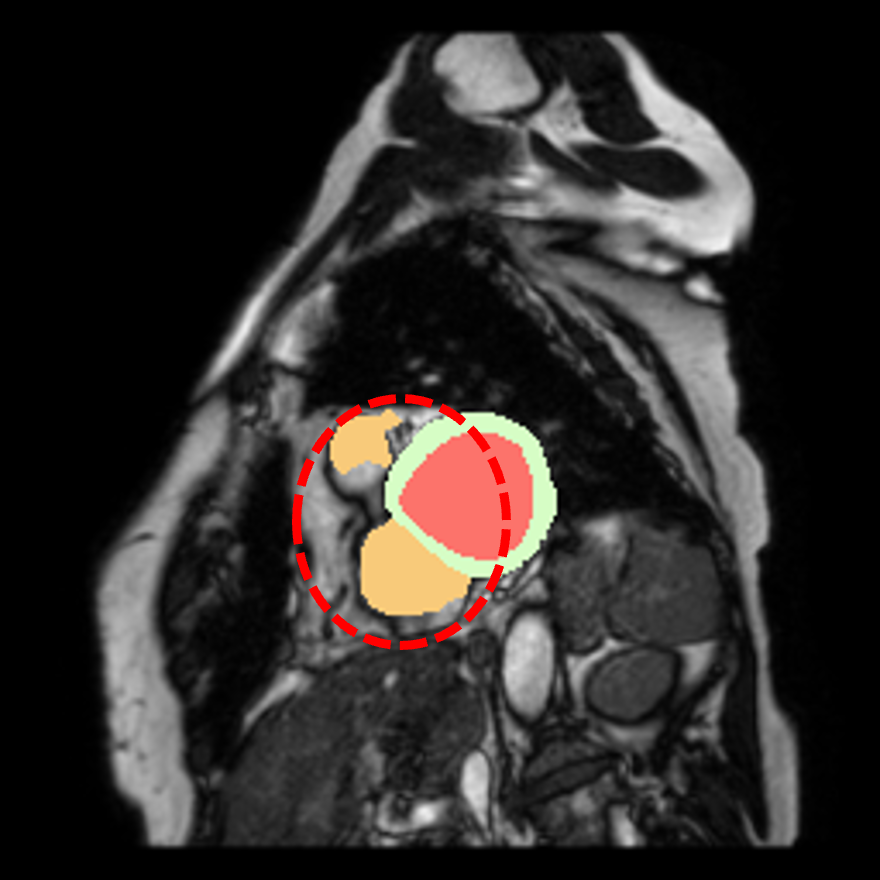}\vspace{3pt}
\includegraphics[width=1\linewidth]{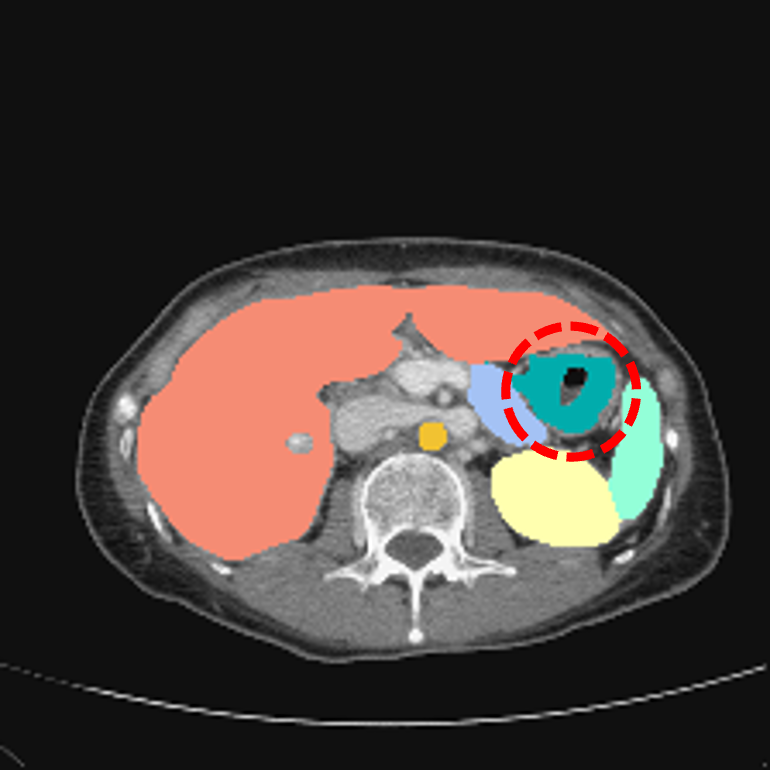}\vspace{3pt}
\includegraphics[width=1\linewidth]{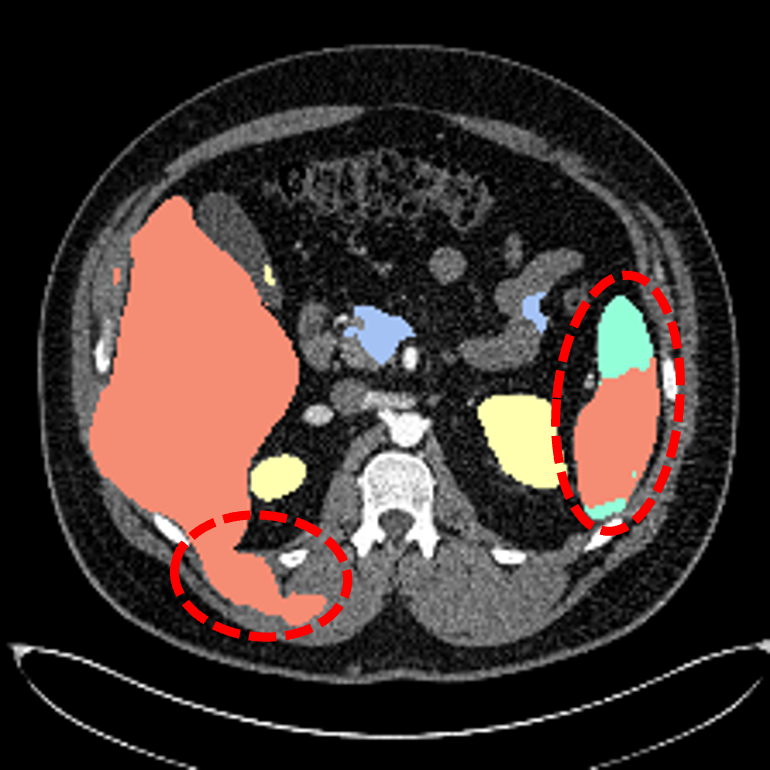}
\end{minipage}}\hspace{-1.3mm}
\subfigure[\our]{
\begin{minipage}[b]{0.235\linewidth}
\includegraphics[width=1\linewidth]{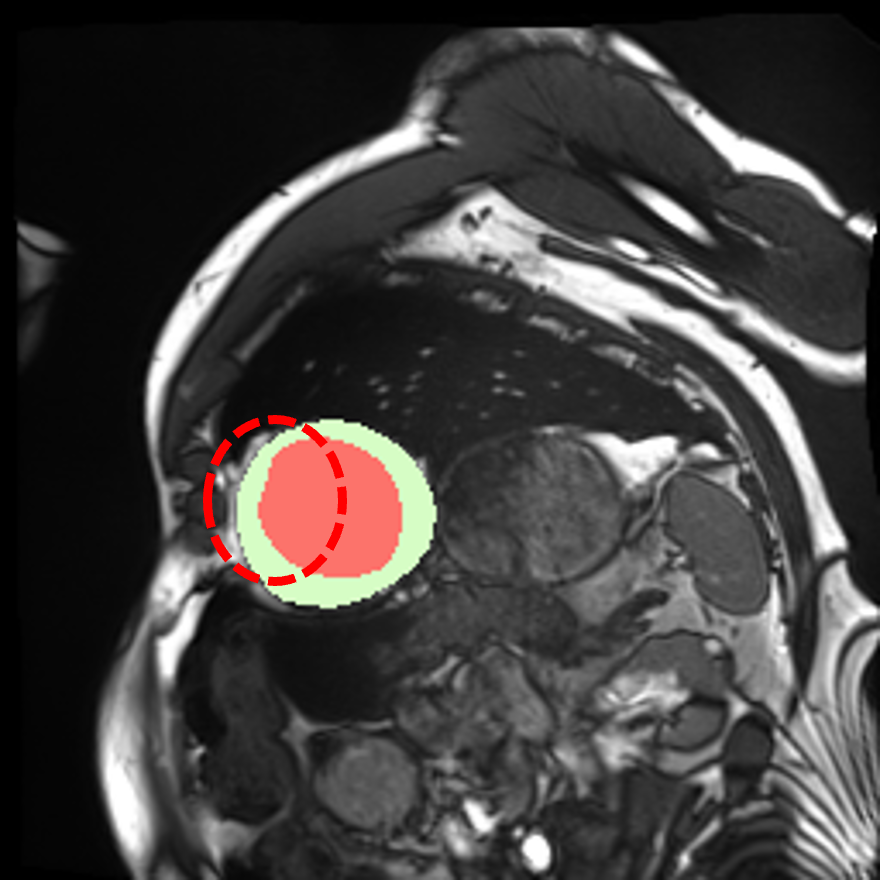}\vspace{3pt}
\includegraphics[width=1\linewidth]{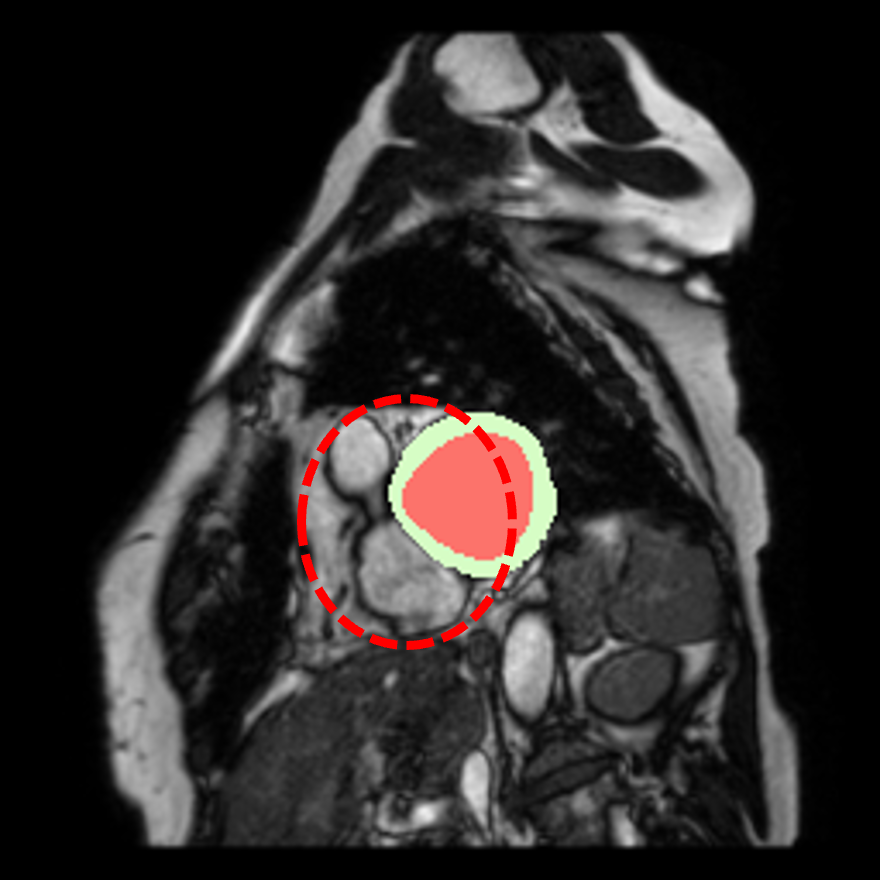}\vspace{3pt}
\includegraphics[width=1\linewidth]{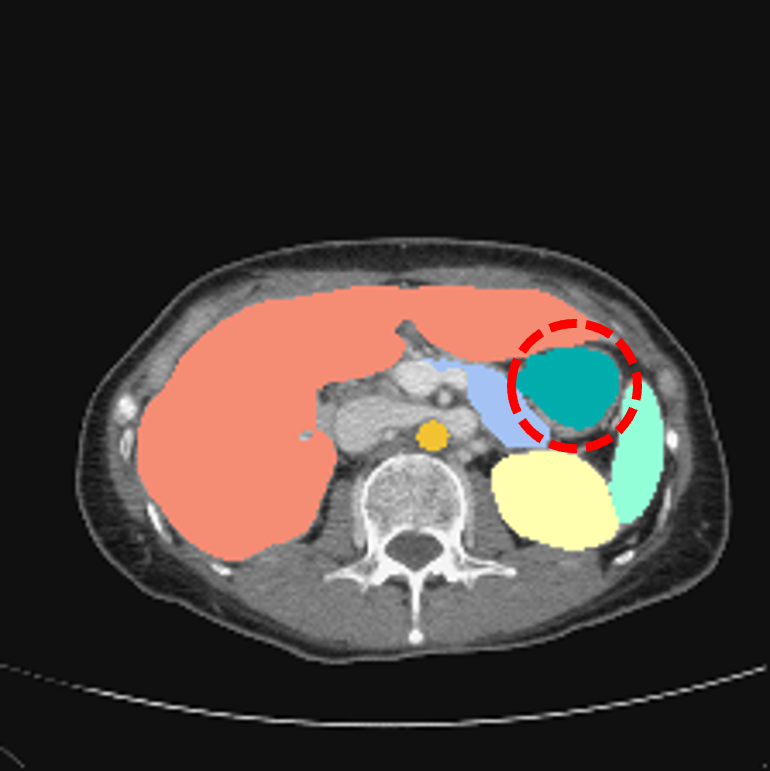}\vspace{3pt}
\includegraphics[width=1\linewidth]{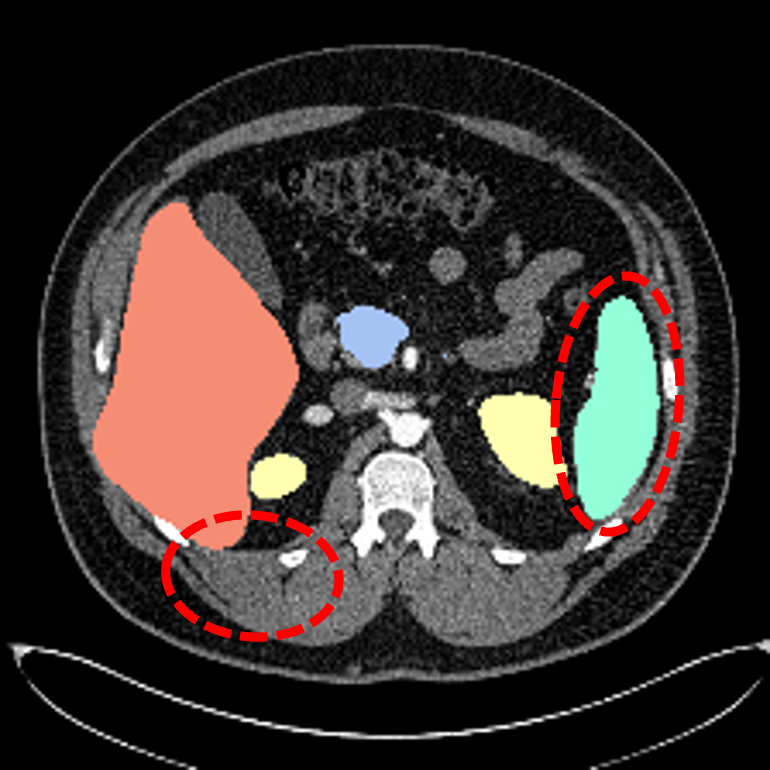}
\end{minipage}}\hspace{-1.3mm}
\subfigure[Ground truth]{
\begin{minipage}[b]{0.235\linewidth}
\includegraphics[width=1\linewidth]{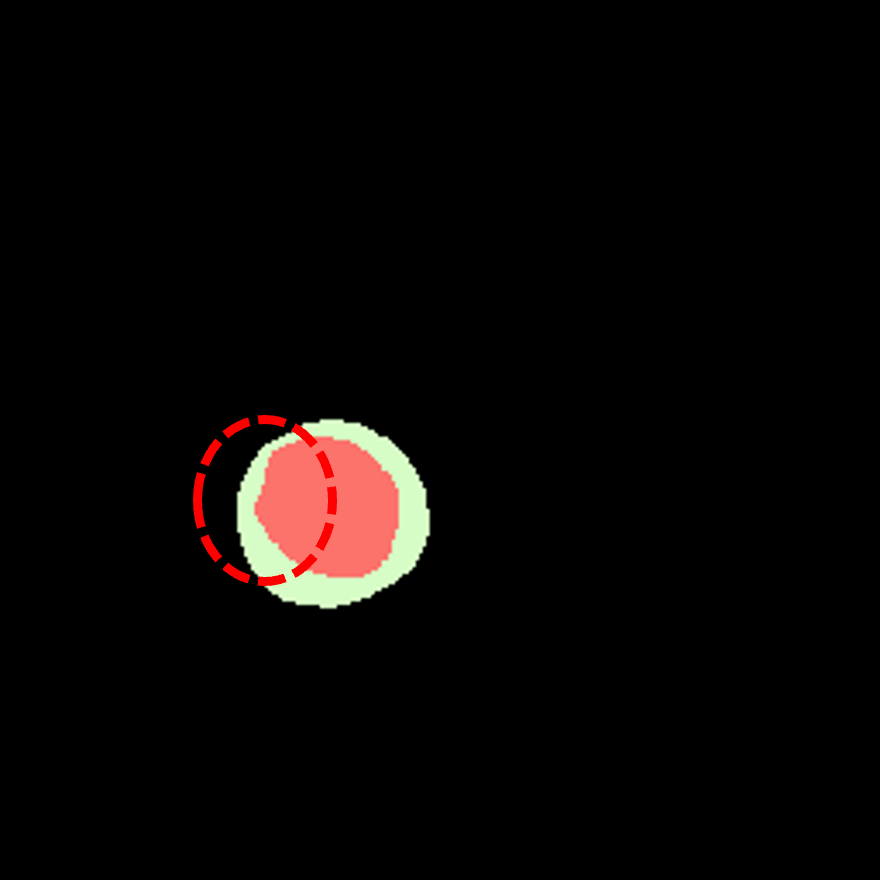}\vspace{3pt}
\includegraphics[width=1\linewidth]{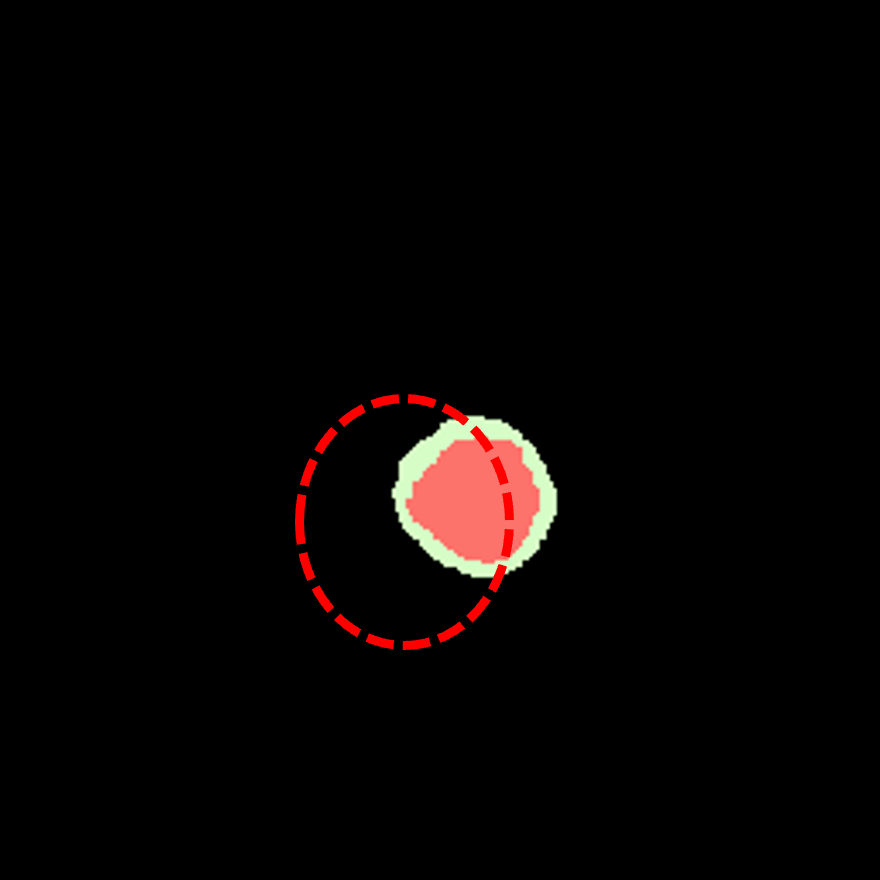}\vspace{3pt}
\includegraphics[width=1\linewidth]{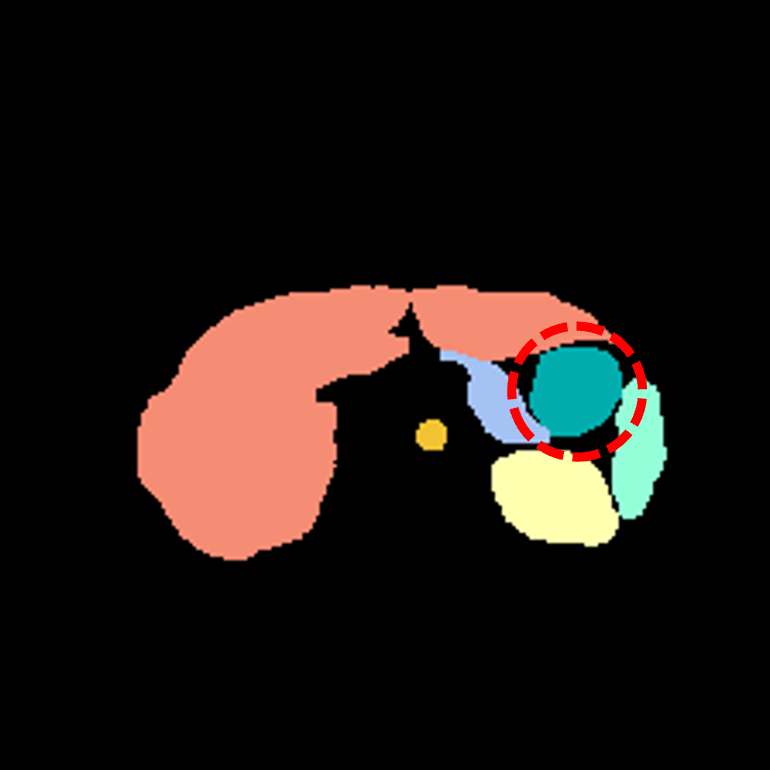}\vspace{3pt}
\includegraphics[width=1\linewidth]{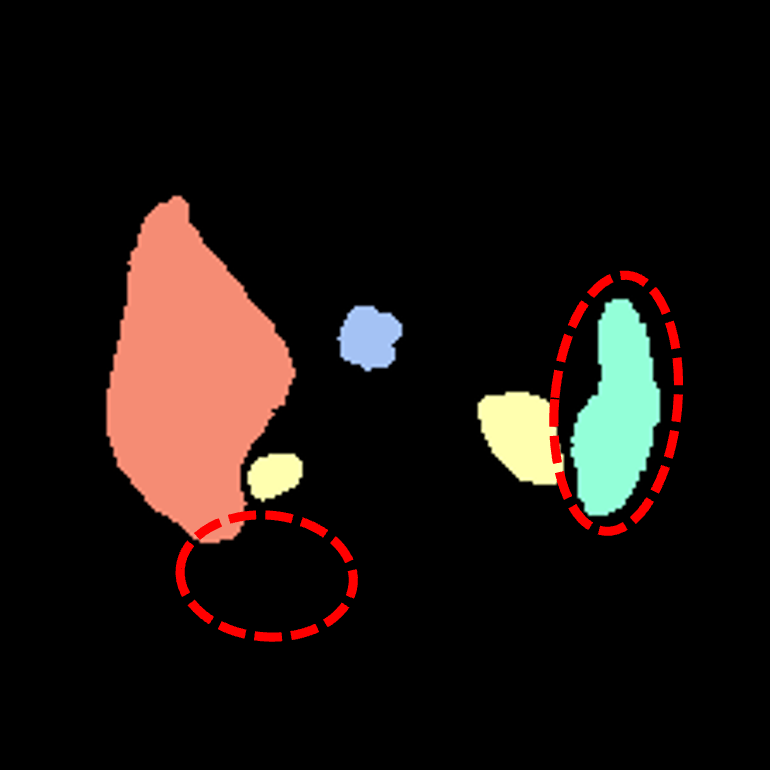}
\end{minipage}}
\caption{Some results of the proposed \our built
  upon the baseline U-Net on the cardiac segmentation (Top two rows: \revise{intra-} and cross-dataset) and some qualitative illustration of \our built on the baseline TransUNet for multi-organ segmentation (Bottom two rows: \revise{intra-} and cross-dataset).
  }
\label{fig:ACDC-Synapse}
\end{figure}
%aining epochs on the test set of Kvasir and CVC-ClinicDB.

% \begin{figure}[ht]
% \centering
% \includegraphics[width=\linewidth]{figure/ACDC-Synapse.png}
% %\vspace{\fixedvskip}
% %\vspace{-6mm}
% \caption{Some cardiac segmentation results of the proposed \our built
%   upon the baseline U-Net on the cardiac segmentation in MRI images (Top) and some qualitative illustration of \our built on TransUnet for multi-organ segmentation in CT images (Bottom).
%   }
% \label{fig:ACDC-Synapse}
% \end{figure}

\subsection{Implementation details}
\label{subsec:implementation}
Since the major goal of this paper is not to obtain state-of-the-art results on each segmentation task, but to show that the proposed \our is 
an effective anatomy-aware segmentation method by incorporating shape-intensity information, which contributes to robust medical image segmentation.
We simply adopt several open-sourced segmentation networks on each segmentation task, including the most widely used U-Net~\cite{ronneberger2015u} for medical image
segmentation, PraNet~\cite{fan2020pranet}, SANet~\cite{wei2021shallow}, TransUNet~\cite{chen2021transunet}, MaxStyle~\cite{chen2022maxstyle}, \revise{LM-Net~\cite{LU2024107717}, 2D D-LKA Net~\cite{azad2024beyond}} and recent SAMed~\cite{zhang2023customized}, which is based on the vision foundation model SAM~\cite{kirillov2023SAM}.
We also implement the proposed \our on the baseline (BL) of
SAUNet~\cite{sun2020saunet}, one of the very few shape-aware medical
image segmentation methods that release the source code. Note that
both the teacher and student networks have exactly the
same network architecture, except for the U-Net and TransUNet, where
the skip-connection is discarded for the teacher segmentation
network. This is because segmenting the class-wise averaged
training images is a relatively simple task. Using the skip-connection
may prevent the learning of shape-intensity information for the penultimate
layer of the teacher model.  All the experiments are conducted using
the Pytorch framework on a workstation with a NVIDIA GeForce RTX 3090
GPU (24GB).
% In all experiments, the hyper-parameter $\alpha$ involved in Eq.~\eqref{eq:losstotal} is set to 2. 
% We adopt the same training
% settings including some classical data augmentation (such as random
% rotation and geometric affine transformation) for both the baseline
% model and the proposed \our in all experiments.

% All the experiments are conducted using
% the Pytorch framework on a workstation with a NVIDIA GeForce RTX 3090
% GPU (24GB). In all experiments, the hyper-parameter $\alpha$ involved
% in Eq.~\eqref{eq:losstotal} is set to 2.  We adopt the same training
% settings including some classical data augmentation (such as random
% rotation and geometric affine transformation) for both the baseline
% model and the proposed \our in all experiments.

%% We adopt the same hyper-parameters and optimization related parameters
%% (from the open-sourced implementation) for both the baseline model and
%% the proposed \our. For all experiments, We apply some classical data
%% augmentations such as random rotation and geometric affine
%% transformation for all methods.

\subsection{Experimental results}
\label{subsec:experiments}
%To exhibit the convincingness of our proposed method,
%We performed a series of contrast and ablation experiments. The experimental results are detailed as follows:

% We conduct experiments on five medical image segmentation tasks with
% different imagining modalities to demonstrate the effectiveness of the
% proposed Shapekd. We detail the corresponding experimental results in
% the following.

\medskip
\noindent
\textbf{Cardiac segmentation:}
% We first validate the proposed \our on
% automatic cardiac segmentation in MRI images. 
We first simply build
the proposed \our using the \textit{de facto} U-Net for medical image
segmentation. Some qualitative results are illustrated in
Fig.~\ref{fig:ACDC-Synapse}. \our achieves anatomically correct segmentation results, and
alleviates the problem of intra-class inconsistency for the baseline
U-Net under both intra-dataset and cross-dataset settings. The quantitative comparison is given in Tab.~\ref{tab:ACDC_2}. \our outperforms the corresponding baseline model.
% In particular, \our outperforms the baseline model in
% segmenting all regions. 
% We also implement \our on the baseline of
% SAUNet~\cite{sun2020saunet}. As shown in Tab.~\ref{tab:ACDC_2}, \our
% outperforms both SAUNet and its baseline on ACDC dataset.

We then benchmark the generalization ability of different methods for
automatic cardiac segmentation in MRI images. As depicted in
Tab.~\ref{tab:ACDC_2}, when generalizing the models trained on the
ACDC dataset to M\&Ms dataset, \our built on U-Net boosts the
performance of the baseline model by 9.9\% Dice score and 1.13 mm HD.
\our also shows improvements over SAUNet and U-Net w/ SDM~\cite{xue2020shape} under this
cross-dataset evaluation. 
Besides, building our SIKD upon \revise{recent MaxStyle~\cite{chen2022maxstyle}, SAMed~\cite{zhang2023customized}, LM-Net~\cite{LU2024107717} and 2D D-LKA Net~\cite{azad2024beyond}} with strong generalization capabilities further consistently enhances their cross-dataset segmentation performance.
The quantitative benchmark on cardiac segmentation confirms that \our is
effective in incorporating the shape-intensity information, thus enhancing the generalization ability of medical image segmentation.
%% Finally, we also evaluate different models
%% on the corrupted ACDC dataset~\cite{chen2021cooperative}, \our
%% achieves consistent or significant improvement over the baseline
%% models in terms of Dice score, is performs favorably in terms of HD.
%% The quantitative comparison is given in
%% Tab.~\ref{tab:ACDC_2}. Though the baseline model already achieves
%% very compelling result, the proposed \our still improves the baseline
%% U-Net by 0.79\% Dice score and 7.01 mm HD. Notably, the most
%% significant improvement is on the segmentation of left ventricle (LV)
%% having approximately circle shape. This further demonstrates that the
%% proposed \our is effective in incorporating the shape information to
%% improve the segmentation performance.

%% Although the baseline results are already
%% very excellent, our method can still improve them further. In
%% particular, the results of the evaluation indicator hausdorff distance
%% is reduced by 7.01 mm. As can be observed from the visualization
%% results in Figure \ref{fig:ACDC}, our method more accurately predicts
%% the location of the edges of the myocardium and right ventricular
%% cavity, and more completely predicts the left ventricular cavity, thus
%% validating that shape information helps improve the network's
%% prediction of anatomical morphology.

\begin{table}[h]
	\centering
	% \vspace{\fixedvskip}
	\caption{Comparison on the Synapse multi-organ segmentation dataset~\cite{landman2015miccai} and the AbdomenCT-1K dataset~\cite{ma2021abdomenct}. 
		%$\uparrow$ (\resp $\downarrow$) indicates the higher (\resp lower) the better.
	  % $\dag$ represents the reproduced results based on the open-sourced implementation.
	}

	\setlength{\tabcolsep}{3mm}{
        \footnotesize
		\label{tab:synapse}
		\begin{tabular}{lcccc}
			\toprule
			\multirow{3}{*}{Method} &
			\multicolumn{2}{c}{Intra-dataset} &
			\multicolumn{2}{c}{Cross-dataset}\\
   			\cmidrule(r){2-3}
                \cmidrule(r){4-5}
			& 
                \multicolumn{2}{c}{Synapse} &
			\multicolumn{2}{c}{AbdomenCT-1K}\\
                \cmidrule(r){2-3}
                \cmidrule(r){4-5}
			&DICE$\uparrow$ &HD$\downarrow$  &DICE &HD   \\
			\midrule
                U-Net~\cite{ronneberger2015u} &0.760 &43.64 &0.468 &94.32 \\
                U-Net w/ SDM~\cite{xue2020shape} &0.392 &100.35 &0.257 &152.11\\
			{U-Net + \our} &0.780 &30.05 &0.487 &94.23\\
                \midrule
                Baseline (\BL) in~\cite{sun2020saunet} &0.782 &32.47 &0.556 &92.61 \\
                SAUNet~\cite{sun2020saunet}&0.790 &31.38 &0.598 &86.34 \\
                \BL + \our &0.792 &24.61 &0.670 &74.13 \\
                \midrule
                SwinUnet~\cite{cao2023swin} & 0.791 & 21.55 &- &- \\
                {TransUNet~\cite{chen2021transunet}} &0.775 &27.70 &0.695 &75.97 \\
                {TransUNet + \our} &0.801 &19.66 &0.715 &71.48 \\
                \midrule
                MaxStyle~\cite{chen2022maxstyle} & 0.757 & 26.33 & 0.664 & 52.37\\
                MaxStyle + \our & 0.771 & 21.76 & 0.691 & 43.15\\
                \midrule
                SAMed~\cite{zhang2023customized} & 0.816 & 21.78 & 0.803 & 44.66 \\
                SAMed + \our & 0.816 & 20.19 & 0.819 & 25.99 \\
                \midrule
            \revise{LM-Net~\cite{LU2024107717}} &0.793 & 23.64  &0.727 &50.85 \\
            \revise{LM-Net + \our} & 0.808 & 20.76 &0.755  &38.59\\
            \midrule
            \revise{2D D-LKA Net~\cite{azad2024beyond}} &0.833 &18.96 & 0.797 & 35.56 \\
            \revise{2D D-LKA + \our} & \textbf{0.843} &\textbf{17.29} & \textbf{0.822} &\textbf{22.11} \\
                
			\bottomrule
		\end{tabular}}
\end{table}

%polyp
\begin{table*}[tp]
  \centering
%  \normalsize
  \caption{Quantitative evaluation of polyp segmentation on Kvasir~\cite{jha2020kvasir}, CVC-ClinicDB~\cite{bernal2015wm}, CVC-ColonDB~\cite{tajbakhsh2015automated}, ETIS~\cite{silva2014toward}, and CVC-T~\cite{vazquez2017benchmark} dataset.
    %$\uparrow$ (\resp $\downarrow$) indicates the higher (\resp lower) the better.
    % $\dag$ represents the reproduced results based on the open-sourced implementation. The other results are from the corresponding original papers.
  }
%   \vspace{\fixedvskip}
  \setlength{\tabcolsep}{3.8mm}{
  \footnotesize%\small
    % \resizebox{\textwidth}{20mm}{
    \label{tab:polyp2}
  \begin{tabular}{lcccccccccc}
    % \hline
    \toprule

    \multirow{4}{*}{Method} &
    \multicolumn{4}{c}{Intra-dataset} &
    \multicolumn{6}{c}{Cross-dataset}\\
    \cmidrule(r){2-5}
    \cmidrule(l){6-11}
    &
    \multicolumn{2}{c}{Kvasir} &
    \multicolumn{2}{c}{CVC-ClinicDB}&
     \multicolumn{2}{c}{CVC-ColonDB}&
    \multicolumn{2}{c}{ETIS}&
     \multicolumn{2}{c}{CVC-T}
    \\
    \cmidrule(r){2-3}
    \cmidrule(lr){4-5}
    \cmidrule(lr){6-7}
    \cmidrule(lr){8-9}
    \cmidrule(l){10-11}
      &   IoU$\uparrow$ &DICE$\uparrow$ &IoU &DICE &IoU &DICE &IoU &DICE &IoU &DICE \\
    % \hline
    \midrule
    SFA~\cite{fang2019selective}  &0.611 &0.723 &0.607 &0.700 &0.347 &0.469 &0.217 &0.297 &0.329 &0.467 \\
    U-Net++~\cite{zhou2018unet++}  &0.743 &0.821  &0.729 &0.794 &0.410 &0.483 &0.344 &0.401 &0.624 &0.707 \\
    PraNet~\cite{fan2020pranet}  &0.840 &0.898 &0.849 &0.899 &0.640 &0.709 &0.567 &0.628 &0.797 &0.871 \\
    SANet~\cite{wei2021shallow}  &0.847  &0.904 &0.859 &0.916 &0.670 &0.753 &0.654 &0.750 &0.815 &0.888\\
    LDNet~\cite{zhang.LDNet} &0.853 &0.907 &\textbf{0.895} &\textbf{0.943} &0.706 &0.784 &0.665 &0.744 &- &-\\
    \midrule
    U-Net~\cite{ronneberger2015u}  &0.761 &0.841 &0.838 &0.885 &0.552 &0.598 &0.323 &0.383 &0.697 &0.769\\
    U-Net w/ SDM~\cite{xue2020shape} & 0.766 & 0.849 & 0.843 & 0.895 & 0.567 & 0.641 & 0.349 & 0.405 & 0.630 & 0.706\\
    U-Net + \our  &0.769 &0.851 &0.851 &0.903 &0.576 &0.650 &0.445 &0.513  &0.712 &0.788\\
    \midrule
    Baseline (\BL) in~\cite{sun2020saunet} &0.810 &0.867 &0.829 & 0.886 &0.606 &0.679 &0.448 &0.510 &0.780 &0.849   \\
    SAUNet~\cite{sun2020saunet} &0.812 &0.870 &0.825 & 0.880 &0.587 &0.658 &0.536 &0.607 &0.761 &0.830 \\
    \BL + \our &0.836 &0.889 &0.850 &0.896 &0.619 &0.687 &0.532 &0.600 &0.791 &0.866 \\
    \midrule
    PraNet~\cite{fan2020pranet}  &0.834 &0.892 &0.859 &0.909 &0.649 &0.721 &0.579 &0.653 &0.818 &0.891\\
    PraNet + \our  &0.852 &0.904 &0.875 &0.927 &0.657 &0.733 &0.607 &0.679 &\textbf{0.830} &\textbf{0.897} \\
    \midrule
    SANet~\cite{wei2021shallow}  &0.845  & 0.903 &0.861 &0.916 &0.679 &0.762 &0.642 &0.741 &0.785 &0.870\\
    SANet + \our &\textbf{0.856}  &\textbf{0.909} &0.880 &0.929 &\textbf{0.712} &\textbf{0.790} &\textbf{0.678} &\textbf{0.760} &0.823 &0.892\\
    \midrule
    MaxStyle~\cite{chen2022maxstyle} & 0.724 & 0.808 & 0.724 & 0.790 & 0.493 & 0.593 & 0.305 & 0.364 & 0.657 & 0.769\\
    MaxStyle + SIKD & 0.755 & 0.835 & 0.747 & 0.813 & 0.510 & 0.603 & 0.361 & 0.427 & 0.671 & 0.780\\
    \midrule
    SAMed~\cite{zhang2023customized} & 0.836 & 0.896 & 0.795 & 0.865 & 0.665 & 0.742 & 0.565 & 0.644 & 0.763 & 0.841\\
    SAMed + \our & 0.838 & 0.897 & 0.802 & 0.872 & 0.688 & 0.761 & 0.626 & 0.709 & 0.786 & 0.859\\
\midrule
\revise{LM-Net~\cite{LU2024107717}} &0.827 &0.893 &0.817 &0.872 &0.653 &0.734 &0.565 &0.645 &0.752 &0.835 \\
\revise{LM-Net + \our} &0.835 &0.895 &0.844 &0.896 &0.687 &0.765 &0.598 &0.689 &0.790 &0.864 \\
\midrule
\revise{2D D-LKA Net~\cite{azad2024beyond}}  &0.833 &0.890 &0.823 &0.878 &0.664 &0.742 &0.523 &0.599 &0.771 &0.832 \\
\revise{2D D-LKA + \our}  &0.846 &0.902 &0.831 &0.885 &0.702 &0.779 &0.606 &0.678 &0.805 &0.876 \\

    \bottomrule

  \end{tabular}
  }
\end{table*}

\textbf{Multi-organ segmentation:} 
For multi-organ segmentation in CT images, 
% following \cite{chen2021transunet, cao2023swin}, we take each
% slice of the 3D volume as an input to the network and stack the
% predicted 2D slices together to reconstruct the 3D prediction for
% evaluation.
Some illustrative results are shown in
Fig.~\ref{fig:ACDC-Synapse}. 
Qualitatively, \our built on TransUNet~\cite{chen2021transunet} accurately segments different
organs and preserves their shapes well.  
The quantitative benchmark on the Synapse multi-organ segmentation dataset is depicted in Tab.~\ref{tab:synapse}. 
\begin{figure}[h!]
\centering
\includegraphics[width=\linewidth]{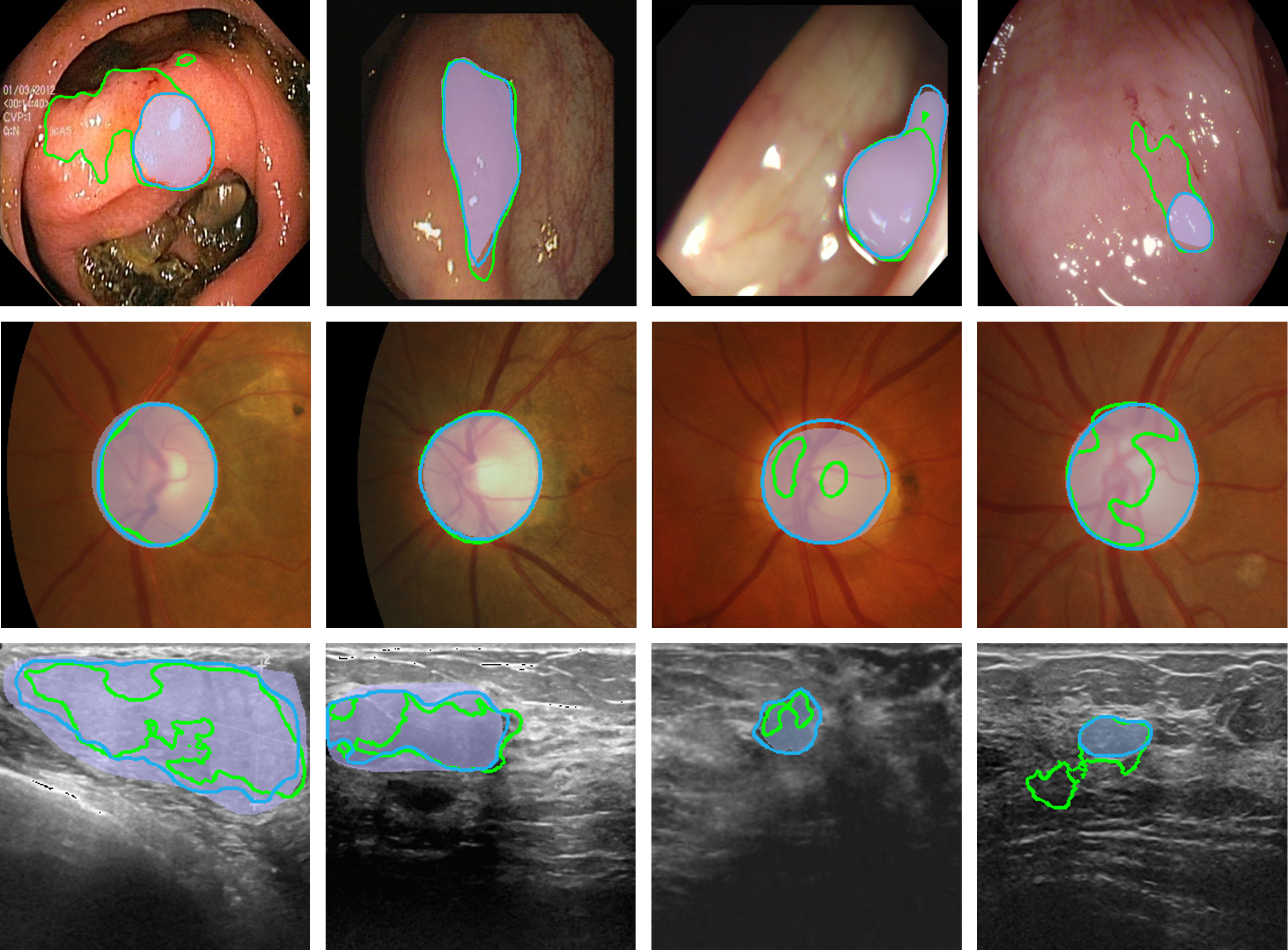}
%\vspace{\fixedvskip}
%\vspace{-6mm}
\caption{Some results on the intra-dataset (left two columns) and cross-dataset (right two columns) of polyp segmentation, ONH segmentation, and breast tumor segmentation (from top to bottom). Green outline: segmentation by the baseline U-Net model; Blue outline: segmentation by \our built upon U-Net; Light blue area: ground-truth segmentation.}
\label{fig:three-dataset}
\end{figure}
\our performs better than the corresponding baseline in terms of both  Dice score and HD, implying that \our achieves good surface prediction and preserves the shapes better.  
Specifically, \our outperforms the baseline TransUNet by 2.69\% Dice score and 8.04 mm HD. Compared with SwinUnet~\cite{cao2023swin}, \our based on TransUNet achieves an improvement of 1.01\% Dice score and 1.89 mm HD. Implementing \our with MaxStyle~\cite{chen2022maxstyle} and SAMed~\cite{zhang2023customized} also consistently boosts the intra-dataset segmentation performance.
\revise{The latest 2D D-LKA achieved state-of-the-art (SOTA) performance. Based on this approach, our method improves the Dice score by more than 1\%.}
% We also benchmark the generalization ability of different methods by
% conducting cross-dataset evaluation. For that, we also report the
% results on AbdomenCT-1K dataset~\cite{ma2021abdomenct} for the models
% trained on Synapse multi-organ segmentation
% dataset~\cite{landman2015miccai}. 

We also benchmark the generalization ability of different methods by
conducting cross-dataset evaluation on AbdomenCT-1K dataset~\cite{ma2021abdomenct} for the models
trained on Synapse multi-organ segmentation
dataset~\cite{landman2015miccai}. 
As depicted in Tab.~\ref{tab:synapse}, \our outperforms all the
corresponding baseline models under cross-dataset evaluation. In
particular, \our built on the baseline of SAUNet~\cite{sun2020saunet}
improves the baseline model by 11.44\% Dice score and 18.48 mm HD.
Compared with SAUNet which only incorporates shape information, \our
achieves 7.17\% Dice score and 12.21 mm HD improvement. 
Building \our
on SAMed~\cite{zhang2023customized} improves the baseline model by
1.6\% Dice score and 18.67 mm HD. \revise{Based on the SOTA 2D D-LKA method, our approach achieves the best generalization results.}
This demonstrates that \our
effectively incorporates the shape-intensity knowledge and generalizes well to
images of unseen dataset.

% We adopt the
% U-Net~\cite{ronneberger2015u}, the baseline of
% SAUNet~\cite{sun2020saunet}, PraNet~\cite{fan2020pranet}, and
% SANet~\cite{wei2021shallow} as the corresponding baseline model to
% build the proposed \our for polyp segmentation in colonoscopy
% images.

\medskip
\noindent
\textbf{Polyp segmentation:} 
Some qualitative polyp segmentation results are illustrated in Fig.~\ref{fig:three-dataset}. The proposed \our based on U-Net can accurately segment the polyp.
% on the test set of training dataset
% (Kvasir and CVC-ClinicDB) and the polyp in images of unseen datasets
% (CVC-ColonDB, ETIS, and CVC-T).
The quantitative comparison with the baseline models and some
state-of-the-art methods is shown in Tab.~\ref{tab:polyp2}. \our
outperforms all the other methods on Kvasir and CVC-ClinicDB dataset,
on which the models are trained. 
In particular, compared with SAUNet which only incorporates the shape
information, \our is more effective with an improvement ranging from
1.6\% to 2.5\%.
% Specifically, \our built on
% U-Net~\cite{ronneberger2015u} improves the baseline U-Net by 0.8\% and
% 1.3\% IoU (\resp 1.0\% and 1.8\% Dice score) on Kvasir and
% CVC-ClinicDB, respectively. And it achieves higher results than the method~\cite{xue2020shape} applying shape constraints.
% Implementing \our with the baseline in SAUNet~\cite{sun2020saunet}, \our achieves 2.6\% and 2.1\% IoU (\resp 2.2\% and 1.0\% Dice score) improvement on Kvasir and CVC-ClinicDB, respectively. Compared with SAUNet which also incorporates shape information, \our is more effective with an improvement ranging from 1.6\% to 2.5\%.   
% Building \our on the PraNet~\cite{fan2020pranet} brings 1.8\% and 1.6\% IoU (\resp 1.2\% and 1.8\% Dice score) improvement on Kvasir and CVC-ClinicDB, respectively. Implementing
% \our with SANet~\cite{wei2021shallow} further boosts the segmentation performance of SANet by 1.1\% IoU and 0.6\% Dice score on Kvasir dataset, and by 1.9\% IoU and 1.3\% Dice score on CVC-ClinicDB dataset. 
\our built on SANet is comparable with recent LDNet~\cite{zhang.LDNet} which relies on lesion-aware dynamic kernel and cross and self-attention modules.
\our built on MaxStyle~\cite{chen2022maxstyle}, SAMed~\cite{zhang2023customized} \revise{LM-Net~\cite{LU2024107717} and 2D D-LKA Net~\cite{azad2024beyond}} further boosts the segmentation performance of the corresponding baseline method.
It is noteworthy that these improvements are achieved without
any extra runtime and memory cost during inference compared with corresponding baseline methods.

\begin{figure}[tp]
\centering
\includegraphics[width=\linewidth]{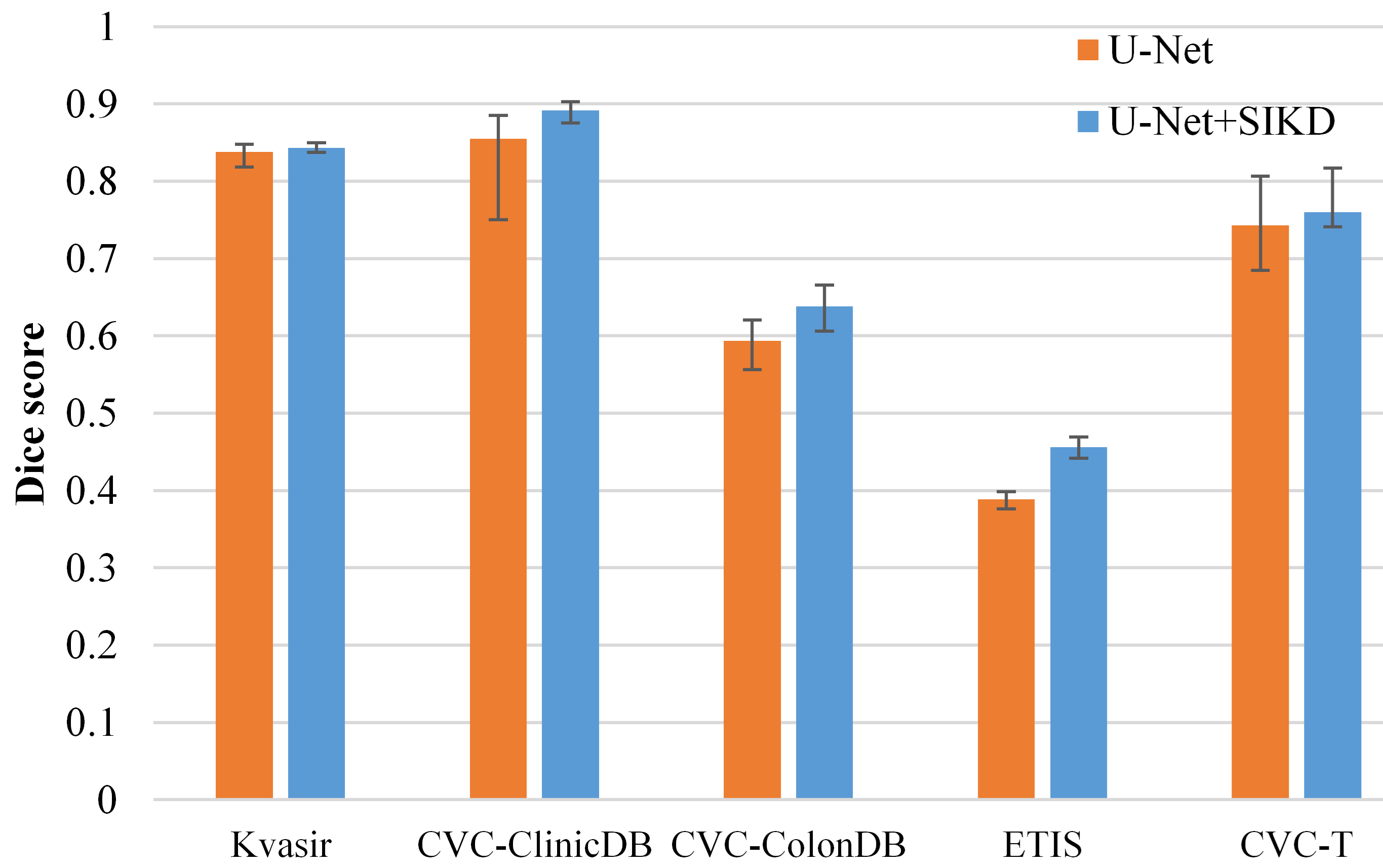}
\caption{Comparison on distribution of dice scores for polyp segmentation when training 10 times the baseline U-Net and the proposed \our on training images of Kvasir~\cite{jha2020kvasir} and CVC-ClinicDB~\cite{bernal2015wm}.}
\label{fig:polyp_histogram}
\end{figure}

We also evaluate the proposed \our on the other three unseen datasets to assess the generalizability of \our
for polyp segmentation. As depicted in Tab.~\ref{tab:polyp2}, using
the U-Net as the baseline model, \our achieves 2.4\%, 12.2\%, and
1.5\% IoU (\resp, 5.2\%, 13.0\%, and 1.9\% Dice score) on CVC-ColonDB,
ETITS, CVC-T dataset, respectively. 
\our based on the baseline of
SAUNet, PraNet, SANet, MaxStyle, SAMed, \revise{LM-Net~\cite{LU2024107717} and 2D D-LKA Net~\cite{azad2024beyond}} also achieves consistent improvement on
these three unseen datasets.
%% Applying \our on SANet \textcolor{blue}{(cite???)} further boosts
%% the segmentation performance by 3.3\% IoU and 2.8\% Dice score,
%% 3.6\% IoU and 1.9\% Dice score, and 3.8\% IoU and 2.2\% Dice score
%% on CVC-ColonDB, ETIS, and CVC-T, respectively.
This demonstrates that the proposed \our generalize well to unseen
datasets. In particular, \our with SANet performs better than the most
recent LDNet~\cite{zhang.LDNet} in generalizing to images of unseen
datasets.

We observe that the performance improvement for segmenting polyp across domains is not as significant as that for the other segmentation tasks (see Fig.~\ref{fig:fig1}). Therefore, we further perform statistical analysis on polyp segmentation by training 10 times the baseline U-Net and the proposed \our built upon U-Net. The statistical results is depicted in Fig.~\ref{fig:polyp_histogram}. 
%experimental results are relatively weakly boosted on cross-domain data. In order to further validate the effectiveness of the proposed method, we conducted multiple training runs (10 times) for the proposed method built on U-Net and the baseline model. 
%The experimental results were then statistically analyzed and presented in the form of a histogram in Fig.~\ref{fig:polyp_histogram}. 
The histogram in Fig.~\ref{fig:polyp_histogram} shows that \our achieves higher average performance and more stable results than the baseline model.

%not only achieves higher average results than the baseline, and is also more stable. 

%but also exhibits smaller fluctuations.
%% Compared with the SANet, \our built on PraNet
%% performs worse on CVC-ColonDB and ETIS, and better on CVC-T. Indeed,
%% SANet is a stronger baseline than PraNet. Besides, SANet leverages
%% color exchange for data augmentation and probability correction
%% strategy for post-processing, which may also improve \our build on
%% PraNet. In fact, applying \our on SANet would probably further boost
%% the performance.

% Considering that different seed and different initialization of U-Net
% usually lead to different segmentation results, we trained the
% baseline U-Net and the proposed \our 10 times under the same setting,
% and compare the overall segmentation performances. The distribution of
% Dice scores is shown in Tab.~\ref{fig:polyp_histogram}. In general, \our
% achieves better performance and is more stable.

\begin{table}[tp]
	\centering
	%   \normalsize
	\caption{Quantitative result on optic nerve head segmentation in fundus images. The models are trained on training set of REFUGE~\cite{orlando2020refuge}. 
 % \dag represents the reproduced results based on the open-sourced implementation. %Red (bold) denotes the best result in each group (column).
          %$\uparrow$ (\resp $\downarrow$) indicates the higher (\resp lower) the better.
        }
	\label{tab:Fundus}
	% \vspace{\fixedvskip}
	% \setlength{\tabcolsep}{3pt}
	\setlength{\tabcolsep}{1.5mm}{
         \footnotesize
		\begin{tabular}{lcccccc}
			\toprule
			\multirow{4}{*}{Method} &
			\multicolumn{2}{c}{Intra-dataset} &
			\multicolumn{4}{c}{Cross-dataset}\\
			\cmidrule(r){2-3}
                \cmidrule(l){4-7}
			&
			\multicolumn{2}{c}{REFUGE} &
			\multicolumn{2}{c}{RIM-ONE-r3} &
			\multicolumn{2}{c}{Drishti-GS}\\
			\cmidrule(r){2-3}
                \cmidrule(lr){4-5}
                \cmidrule(l){6-7}
			&   IoU$\uparrow$ &DICE$\uparrow$ &IoU   &DICE   &IoU   &DICE \\
			\midrule
U-Net~\cite{ronneberger2015u} &0.924 &0.960 &0.266 &0.317 &0.433 &0.505\\
U-Net w/ SDM~\cite{xue2020shape} & 0.925 & 0.960 & 0.346 & 0.419 & 0.497 & 0.624\\
U-Net + \our &0.928 &0.962 &0.501 &0.595 &0.627 &0.692\\
\midrule
Baseline (\BL) in~\cite{sun2020saunet} &0.884 & 0.938 &0.696 &0.812 &0.782 &0.874\\
SAUNet~\cite{sun2020saunet} &0.906 &0.950 &0.662 &0.766 &0.840 &0.911 \\
\BL + \our &0.920 &0.958 &0.708 &0.821 &0.861 &0.924 \\
\midrule
MaxStyle~\cite{chen2022maxstyle} & 0.927 & 0.961 & 0.736 & 0.846 & 0.889 & 0.939\\
MaxStyle + \our & \textbf{0.935} & \textbf{0.966} & 0.748 & 0.852 & 0.897 & 0.945 \\
\midrule
SAMed~\cite{zhang2023customized} & 0.920 & 0.958 & 0.772 & 0.870 & 0.906 & 0.950\\
SAMed + \our & 0.922 & 0.960 & \textbf{0.782} & \textbf{0.879} & 0.916 & 0.958 \\

\midrule
\revise{LM-Net~\cite{LU2024107717}} &0.909 &0.952 &0.730 &0.841 &0.933 &0.965\\
\revise{LM-Net + \our}  &0.926 &0.961 &0.777 &0.873 &\textbf{0.939} &\textbf{0.967}\\
\midrule
\revise{2D D-LKA Net~\cite{azad2024beyond}}   &0.897 &0.946 &0.744 &0.851 &0.930 &0.963\\
\revise{2D D-LKA + \our}   &0.918 &0.955 &0.768 &0.869 &0.938 &0.966\\
			\bottomrule
		\end{tabular}
		% \vspace{\fixedvskip}
	}
	%\vspace{2mm}
\end{table}

\medskip
\noindent
\textbf{Optic nerve head segmentation:} We then conduct experiments on
segmenting optic nerve head (ONH) in color fundus images. Some qualitative
segmentation results are shown in Fig.~\ref{fig:three-dataset}. Both the
baseline model and the proposed \our accurately segment ONH on testing
images of the REFUGE dataset, on whose training set the models are
trained. Yet, the baseline model performs poorly on images with slightly different
appearance. This demonstrates that the proposed \our can effectively
incorporate the shape-intensity information, helping to yield accurate segmentation
across domains. 

The quantitative comparison between the baseline U-Net model and the
proposed \our is depicted in Tab.~\ref{tab:Fundus}. On the REFUGE test
set, \our slightly outperforms the baseline model. On the unseen RIM-ONE-r3 and Drishti-GS dataset,
the proposed \our significantly improves the baseline
model. Precisely, \our improves the baseline U-Net model by 23.5\%
(\resp 19.4\%) IoU and 27.8\% (\resp 18.7\%) Dice score on the RIM-ONE-r3
(\resp Drishti-GS) dataset. Similar performance behavior is observed
on \our built on the baseline of SAUNet~\cite{sun2020saunet}. 
It is noteworthy that
though SAMed~\cite{zhang2023customized}, \revise{LM-Net~\cite{LU2024107717} and 2D D-LKA Net~\cite{azad2024beyond} already exhibits a powerful generalization ability, our \our can still boosts the cross-dataset segmentation performance.}
% In particular, \our improves the baseline in SAUNet by 3.6\% IoU and
% 2.0\% Dice score on REFUGE dataset. For the generalization to
% Drishti-GS dataset, \our improve the corresponding baseline by 7.9\%
% IoU and 5.0\% Dice score. \our also consistently outperforms SAUNet
% under both intra-dataset and cross-dataset evaluation.

\begin{table}[tp]
	\centering
	%   \normalsize
	\caption{Quantitative evaluation of different methods for breast tumor segmentation on BUSI~\cite{al2020dataset}, IDFAHSTU~\cite{zhuang2019rdau}, and dataset B~\cite{yap.datasetB}.} 
	\label{tab:BUSI}
	% \vspace{\fixedvskip}
	% \setlength{\tabcolsep}{3pt}
	\setlength{\tabcolsep}{1.5mm}{
         \footnotesize%\small
		\begin{tabular}{lcccccc}
			\toprule
			\multirow{2}{*}{Method} &
			\multicolumn{2}{c}{Intra-dataset}&
			\multicolumn{4}{c}{Cross-dataset} \\
			\cmidrule(r){2-3}
                \cmidrule(l){4-7}
			& 
			\multicolumn{2}{c}{BUSI}&
			\multicolumn{2}{c}{IDFAHSTU}&
			\multicolumn{2}{c}{Dataset B}\\
			\cmidrule(r){2-3}
                \cmidrule(lr){4-5}
                \cmidrule(l){6-7}
			&   IoU$\uparrow$ &DICE$\uparrow$ &IoU  &DICE  &IoU   &DICE  \\
			\midrule
U-Net~\cite{ronneberger2015u} &0.580 &0.676  &0.597 &0.694 &0.397 &0.461\\
U-Net w/ SDM~\cite{xue2020shape} & 0.588 & 0.681 & 0.626 & 0.732 & 0.411 & 0.483\\
U-Net + \our &0.611 &0.709  &0.646 &0.755 &0.449 &0.521\\
\midrule
\BL in~\cite{sun2020saunet} &0.535 &0.642 &0.611 &0.738 &0.362 &0.459\\
SAUNet~\cite{sun2020saunet} &0.598 &0.698 &0.696 &0.805 &0.513 &0.615 \\
\BL + \our &0.622 &0.719 &0.717 &0.824 &0.540 &0.643\\
\midrule
MaxStyle~\cite{chen2022maxstyle} & 0.603 & 0.701 & 0.755 & 0.851 & 0.529 & 0.631\\
MaxStyle + \our & 0.612 & 0.707 & 0.767 & 0.861 & 0.561 & 0.661\\
\midrule
SAMed~\cite{zhang2023customized} & 0.672 & 0.762 & 0.780 & 0.871 & 0.684 & 0.778\\
SAMed + \our & 0.686 & 0.772 & \textbf{0.793} & \textbf{0.879} & \textbf{0.697} & \textbf{0.790}\\

\midrule
\revise{LM-Net~\cite{LU2024107717}} &0.663 &0.755 &0.737 &0.842 &0.571 &0.663\\
\revise{LM-Net + \our}  &0.684 &0.773 &0.764 &0.862 &0.633 &0.719\\
\midrule
\revise{2D D-LKA Net~\cite{azad2024beyond}}   &0.674 &0.762 &0.593 &0.719 &0.671 &0.760\\
\revise{2D D-LKA + \our}   &\textbf{0.691} &\textbf{0.775} &0.655 &0.780 &0.694 &0.781\\
			\bottomrule
		\end{tabular}
		% \vspace{\fixedvskip}
	}
	%\vspace{2mm}
\end{table}

\medskip
\noindent
\textbf{Breast tumor segmentation:} We then conduct experiments on
breast tumor segmentation in ultrasound images. Some qualitative
illustrations are given in Fig.~\ref{fig:three-dataset}. Compared with the
baseline U-Net, \our accurately segments the breast tumor under both
intra-dataset and cross-dataset settings.
The quantitative comparison between the baseline U-Net and \our is
depicted in Tab.~\ref{tab:BUSI}. \our achieves significant improvement
over the baseline U-Net.  Specifically, on the BUSI dataset, \our
improves the baseline U-Net by 3.1\% IoU and 3.3\% Dice
score. 
\our also outperforms U-Net w/ SDM~\cite{xue2020shape} by 2.3\% IoU and 2.8\% Dice score.
Implementing \our with the baseline of
SAUNet~\cite{sun2020saunet} improves the baseline by 8.7\% IoU and
7.7\% Dice score. Besides, the proposed \our also outperforms SAUNet
by 2.4\% IoU and 2.1\% Dice score. 
% It is noteworthy that these
% improvements over the baseline models are achieved without any extra
% memory and runtime cost during inference.

\begin{table*}[tp]
	\centering
	%   \normalsize
	\caption{Evaluation of \our on cardiac and optic nerve head segmentation with different inputs for the teacher network.
		%$\uparrow$ (\resp $\downarrow$) indicates the higher (\resp lower) the better.
	}
	\label{tab:input-ablation}
	% \vspace{\fixedvskip}
	% \setlength{\tabcolsep}{3pt}
	\setlength{\tabcolsep}{3.7mm}{
         \footnotesize%\small
		\begin{tabular}{lcccccccccc}
			\toprule
			\multirow{4}{*}{Input} &
			\multicolumn{4}{c}{Cardiac segmentation in MRI images}&
			\multicolumn{6}{c}{Optic nerve head segmentation in color fundus images}
			 \\
			\cmidrule(r){2-5}
                \cmidrule(l){6-11}
			&
			\multicolumn{2}{c}{ACDC} &
			\multicolumn{2}{c}{M\&Ms}&
			\multicolumn{2}{c}{REFUGE} &
			\multicolumn{2}{c}{RIM-ONE-r3} &
			\multicolumn{2}{c}{Drishti-GS}

			\\
			\cmidrule(r){2-3}
                \cmidrule(lr){4-5}
                \cmidrule(lr){6-7}
                \cmidrule(lr){8-9}
                \cmidrule(l){10-11}
			 &Dice$\uparrow$   &HD$\downarrow$ &Dice &HD &IoU &DICE &IoU &DICE &IoU  &DICE\\
			\midrule
                Baseline U-Net~\cite{ronneberger2015u} &0.884 &1.30 &0.686 &4.08 &0.924 &0.960 &0.266 &0.317 &0.433 &0.505  \\
			Original image &0.882 &1.26 &0.693 &4.09 &0.927 &0.961 &0.371 &0.446 &0.571 &0.631 \\
                Annotated label map & 0.893 &\textbf{1.16} & 0.723 &3.69 & 0.922 &0.961 &0.383 &0.452 &0.552 &0.607 \\
                % \rowcolor{gray!20}
			Class-wise mean average &\textbf{0.894} &1.21 &\textbf{0.785} &\textbf{2.95} &\textbf{0.928} &\textbf{0.962} &\textbf{0.501} &\textbf{0.595} &\textbf{0.627} &\textbf{0.692} \\
            
			\bottomrule
		\end{tabular}
		% \vspace{\fixedvskip}
	}
	%\vspace{2mm}
\end{table*}

\begin{figure}[t]
\centering

\subfigure[Image]{
% minipage中放置4张图，通过vspace进行换行调整，但是都在subfigure下面，共享一个子标题
\begin{minipage}[b]{0.188\linewidth}
\includegraphics[width=1\linewidth]{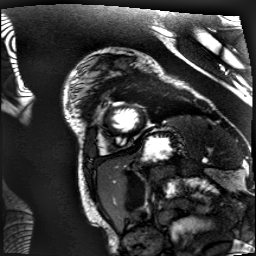}\vspace{3pt}
\includegraphics[width=1\linewidth]{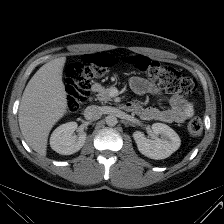}\vspace{3pt}
\includegraphics[width=1\linewidth]{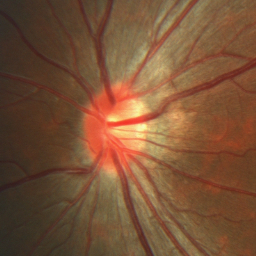}\vspace{3pt}
\includegraphics[width=1\linewidth]{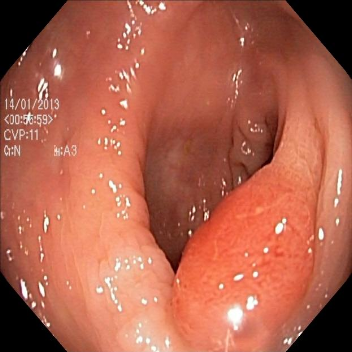}\vspace{3pt}
\includegraphics[width=1\linewidth]{figure/fea_vis/BUSI/FCS-BUSIvisimg-fig7.png}
\end{minipage}}\hspace{-1.3mm}
\subfigure[U-Net]{
\begin{minipage}[b]{0.188\linewidth}
\includegraphics[width=1\linewidth]{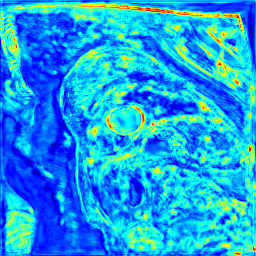}\vspace{3pt}
\includegraphics[width=1\linewidth]{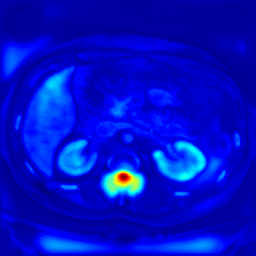}\vspace{3pt}
\includegraphics[width=1\linewidth]{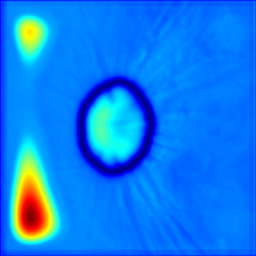}\vspace{3pt}
\includegraphics[width=1\linewidth]{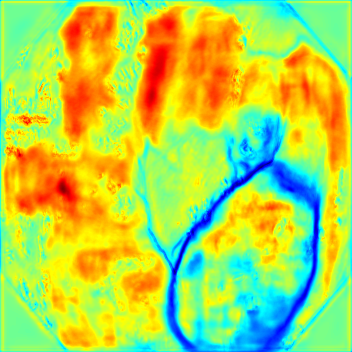}\vspace{3pt}
\includegraphics[width=1\linewidth]{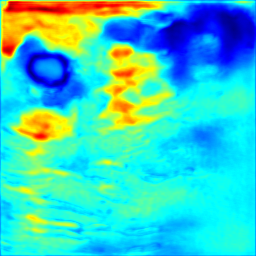}
\end{minipage}}\hspace{-1.3mm}
\subfigure[Teacher]{
\begin{minipage}[b]{0.188\linewidth}
\includegraphics[width=1\linewidth]{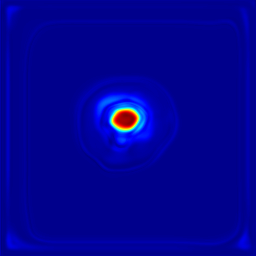}\vspace{3pt}
\includegraphics[width=1\linewidth]{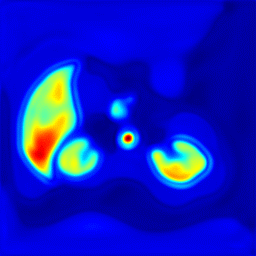}\vspace{3pt}
\includegraphics[width=1\linewidth]{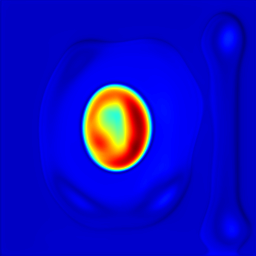}\vspace{3pt}
\includegraphics[width=1\linewidth]{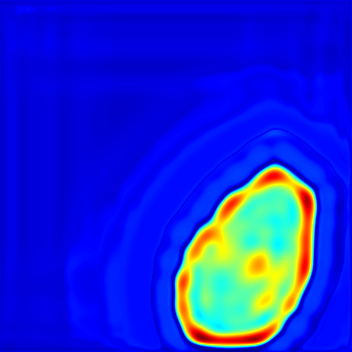}\vspace{3pt}
\includegraphics[width=1\linewidth]{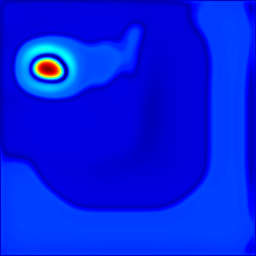}
\end{minipage}}\hspace{-1.3mm}
\subfigure[\our]{
\begin{minipage}[b]{0.188\linewidth}
\includegraphics[width=1\linewidth]{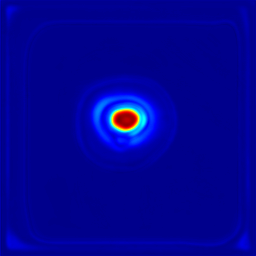}\vspace{3pt}
\includegraphics[width=1\linewidth]{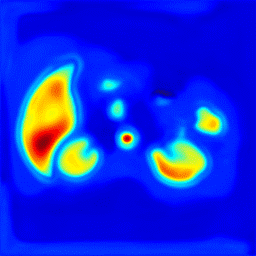}\vspace{3pt}
\includegraphics[width=1\linewidth]{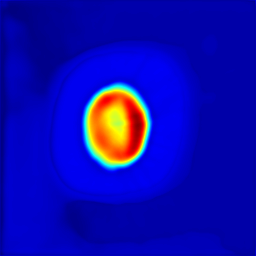}\vspace{3pt}
\includegraphics[width=1\linewidth]{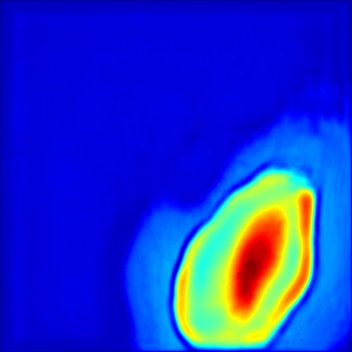}\vspace{3pt}
\includegraphics[width=1\linewidth]{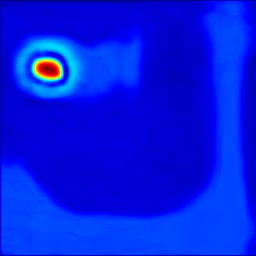}
\end{minipage}}\hspace{-1.3mm}
\subfigure[GT]{
\begin{minipage}[b]{0.188\linewidth}
\includegraphics[width=1\linewidth]{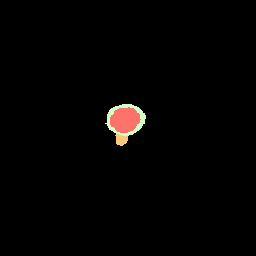}\vspace{3pt}
\includegraphics[width=1\linewidth]{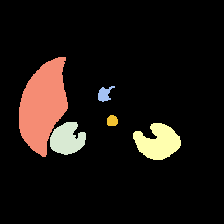}\vspace{3pt}
\includegraphics[width=1\linewidth]{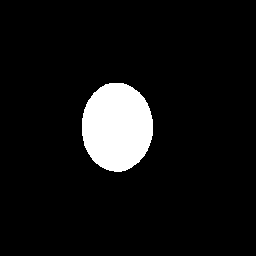}\vspace{3pt}
\includegraphics[width=1\linewidth]{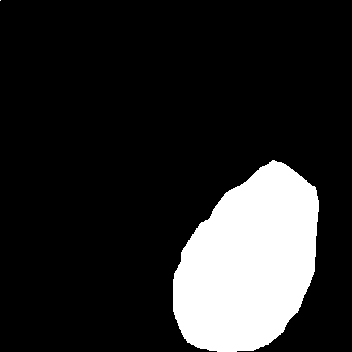}\vspace{3pt}
\includegraphics[width=1\linewidth]{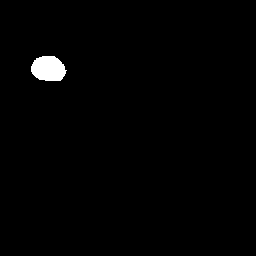}
\end{minipage}}
\caption{Visualization of the penultimate layer feature of baseline U-Net and the proposed \our built upon U-Net.}
\label{fig:fea_vis}
\end{figure}

\begin{table}[tp]
	\centering
	% \tiny
	\caption{Comparison of computational requirements based on U-Net~\cite{ronneberger2015u} architecture. Calculations for training are performed with a batch size of 16 and image dimensions of 256 $\times$ 256 $\times$ 1, while testing is conducted with a batch size of 1 and the same image size.} 
	\label{tab:cost}
	% \vspace{\fixedvskip}
	% \setlength{\tabcolsep}{3pt}
	\setlength{\tabcolsep}{2.3mm}{
         % \footnotesize%\small
		\begin{tabular}{lcccc}
			\toprule
			\multirow{2}{*}{Method} &
			\multicolumn{2}{c}{Training}&
			\multicolumn{2}{c}{Testing} \\
			\cmidrule(r){2-3}
                \cmidrule(l){4-5}
			&FLOPs (G) &GPU (G)   &FLOPs (G) &GPU (G)  \\
			\midrule
U-Net~\cite{ronneberger2015u} &65.46 &9.61  &65.46 &0.25 \\
U-Net + \our &121.21 &10.47  &65.46 &0.25 \\
			\bottomrule
		\end{tabular}
	}
\end{table}

We also evaluate the proposed \our on images from the unseen dataset used
in~\cite{zhuang2019rdau} and the dataset B~\cite{yap.datasetB} to assess the generalizability of \our. As depicted in Tab.~\ref{tab:BUSI}, 
applying \our on the baseline of SAUNet~\cite{sun2020saunet}, MaxStyle~\cite{chen2022maxstyle}, SAMed~\cite{zhang2023customized}, \revise{LM-Net~\cite{LU2024107717} and 2D D-LKA Net~\cite{azad2024beyond}} boosts the segmentation performance of the corresponding baseline.
% using the U-Net as the baseline
% model, \our achieves 4.9\% and 5.2\% IoU (\resp 6.1\% and 6.0\% Dice score) on the
% dataset of~\cite{zhuang2019rdau} and the dataset B~\cite{yap.datasetB}, respectively. Applying \our on the baseline of SAUNet~\cite{sun2020saunet} boosts the segmentation performance of the
% baseline by 10.6\% and 17.8\% IoU (\resp 8.6\% and  18.4\% Dice score) on the
% dataset of~\cite{zhuang2019rdau} and the dataset B~\cite{yap.datasetB}, respectively. \our also outperforms SAUNet by 2.1\% and 2.7\% IoU (\resp 1.9\% and 2.8\% Dice score) on the
% dataset of~\cite{zhuang2019rdau} and the dataset B~\cite{yap.datasetB}, respectively.
This also demonstrates that the proposed \our generalizes well to images of
unseen dataset. It is noteworthy that the performance on the cross-dataset of~\cite{zhuang2019rdau} is better than the intra-dataset performance for breast tumor segmentation.
This is probably because that the dataset in~\cite{zhuang2019rdau}
only contains 42 ultrasound images, which are not as challenging as
the ultrasound images in the BUSI dataset~\cite{al2020dataset} and dataset B~\cite{yap.datasetB}.
% We also evaluate the proposed \our on images from the unseen dataset used
% in~\cite{zhuang2019rdau} to assess the generalizability of \our . As
% depicted in Tab.~\ref{tab:BUSI}, using the U-Net as the baseline
% model, \our achieves 4.9\% IoU and 6.1\% Dice score improvement on the
% dataset of~\cite{zhuang2019rdau}.  Applying \our on the baseline of
% SAUNet~\cite{sun2020saunet} boosts the segmentation performance of the
% baseline by 10.6\% IoU and 8.6\% Dice score. Compared with SAUNet,
% \our outperforms it by 2.1\% IoU and 1.9\% Dice score. This also
% demonstrates that the proposed \our generalize well to images of
% unseen dataset. It is noteworthy that the cross-dataset performance is
% better than the intra-dataset performance for breast tumor segmentation.
% This is probably because that the dataset in~\cite{zhuang2019rdau}
% only contains 42 ultrasound images, which are not as challenging as
% the ultrasound images in the BUSI dataset~\cite{al2020dataset}.

\section{Discussion}
\label{sec:discussion}
% Medical image segmentation plays a crucial role in clinical practice. With the advancement of deep learning, many approaches have achieved good results. However, these methods often regard the segmentation task as an independent pixel classification problem, disregarding the fact that the target in medical images typically has a specific shape, and the intensity prior is also useful for medical image segmentation. Furthermore, some studies~\cite{geirhos2018imagenettrained, li2021shapetexture} have shown that convolutional neural networks often tend to prioritize learning texture information while neglecting shape information. This often leads to anatomically inconsistent segmentation results.

Medical image segmentation plays a crucial role in clinical practice. With the advancement of deep learning, many approaches have achieved notable results. 
% However, these methods often treat segmentation as an independent pixel classification problem, disregarding the fact that targets in medical images typically have specific shapes, and that intensity priors are also valuable for segmentation.
However, many deep-learning-based methods treat segmentation as a pixel-wise classification task, often overlooking the fact that targets in medical images typically possess specific shape-intensity prior information. Additionally, some studies~\cite{geirhos2018imagenettrained, li2021shapetexture} have demonstrated that deep neural networks tend to prioritize learning texture information over shape information. This bias can lead to anatomically implausible segmentation results, especially in cross-dataset segmentation, where domain shifts relative to the training images are present. Therefore, incorporating shape-intensity prior information is beneficial for both intra-dataset and cross-dataset segmentation.

Recently, some methods improve segmentation performance by incorporating shape information, making the segmentation results more reasonable. However, these methods either add edge constraints or further optimize the segmentation results, requiring additional computational resources. Different from existing methods, in this paper, we propose to incorporate joint shape-intensity prior information into the segmentation network. Specifically, we train a teacher network on the class-wise averaged training images to extract the shape-intensity information. We then employ knowledge distillation to transfer the extracted shape-intensity information from the teacher network to the student network (\ie, the segmentation network).

%For a given training image, we first calculate the pixel-wise average values for each class and replace the corresponding class pixel values, obtaining the mean image. We then use the mean image as input to train the teacher network. Since the mean image does not contain any texture information, the teacher network can accurately segment the corresponding targets, effectively encoding the shape-intensity information. Subsequently, we leverage knowledge distillation to transfer the learned the shape-intensity information from the teacher network to the student network. The student network has the same network architecture as the teacher network and takes the original image as input.
%Extensive experiments demonstrate the effectiveness of the proposed method in improving segmentation performance while achieving better generalization.

The proposed \our relies on transferring the shape-intensity knowledge extracted by
the teacher model. Therefore, we analyze the shape-intensity knowledge by visualizing the penultimate layer feature. As shown in Fig.~\ref{fig:fea_vis}, compared with the baseline model, the feature extracted by the teacher model is smoother and has a more complete shape structure. The feature of the student model is very similar to the teacher model. This implies that the proposed \our effectively incorporates the shape-intensity knowledge learned by the teacher model into segmentation. 

As previously mentioned, our method does not require extra runtime/memory cost during inference.  \revise{We} also analyze the training time requirements.
In comparison to the baseline model, our proposed SIKD involves training an additional teacher model on class-wise averaged images to extract shape-intensity prior information. This information is subsequently transferred to the student model via knowledge distillation (KD). During the training of the student network, it is necessary to forward the teacher network to compute the KD loss. 
% As a result, the proposed SIKD requires slightly more than twice the training time (2.33 times for U-Net-based SIKD) and additional memory (approximately 11.1\% extra) compared to the baseline model. However, the training cost is on par with some other methods.
\revise{As shown in Tab.~\ref{tab:cost}, during training, our method's FLOPs are nearly twice as high as the U-Net (as mentioned in Section 4.2 of the manuscript, the skip connections of the U-Net used as the teacher network have been removed, thereby reducing the complexity of the teacher model), and GPU usage is also slightly higher than U-Net. However, during testing, both our method and the U-Net have the same FLOPs and GPU usage, incurring no additional overhead.}

We then conduct two types of ablation studies on the task of cardiac segmentation and optic nerve head segmentation.

\begin{figure*}[th]
\centering
\subfigure[Effect of using different $\alpha$ for cardiac segmentation]{
% minipage中放置4张图，通过vspace进行换行调整，但是都在subfigure下面，共享一个子标题
\begin{minipage}[b]{0.48\linewidth}
\includegraphics[width=0.9\linewidth]{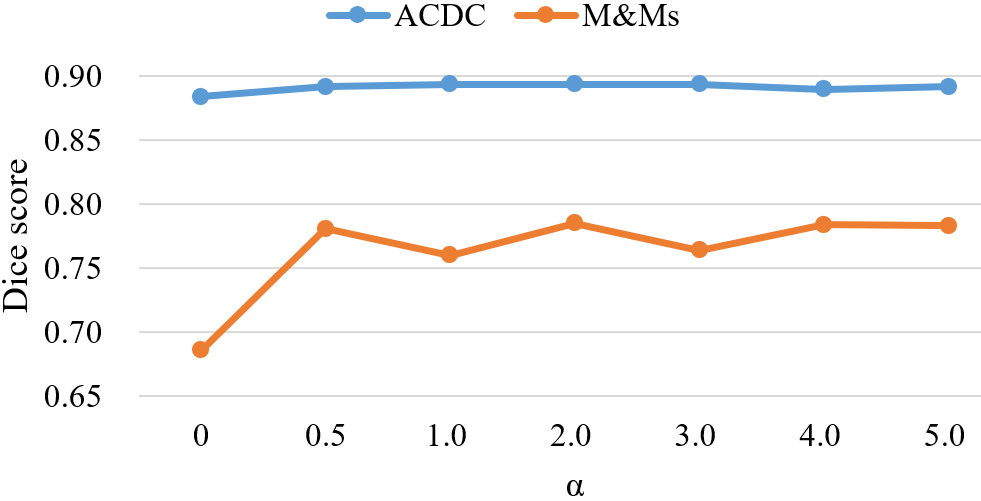}\vspace{3pt}
\end{minipage}}\hspace{-1.3mm}
\subfigure[Effect of using different $\alpha$ for optic nerve head segmentation]{
\begin{minipage}[b]{0.48\linewidth}
\includegraphics[width=0.9\linewidth]{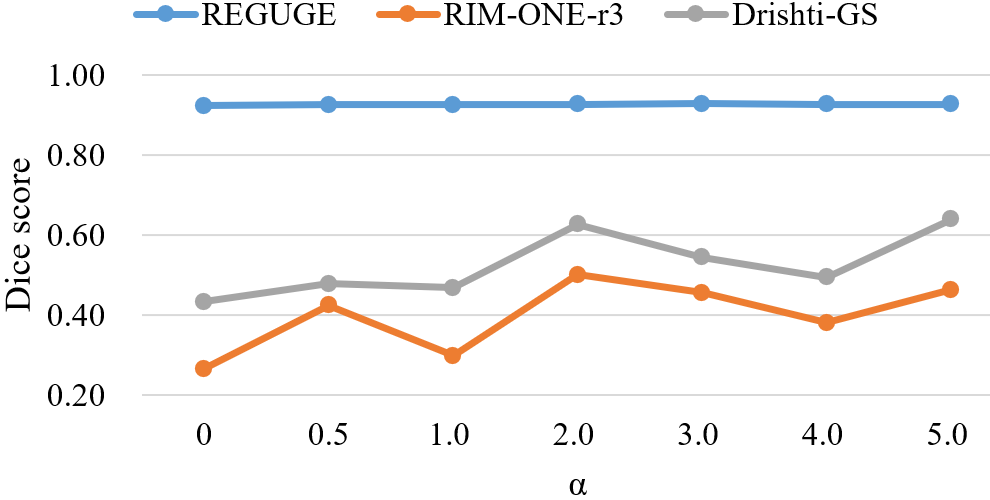}\vspace{3pt}
\end{minipage}}\hspace{-1.3mm}
	\caption{Evaluation of \our with different settings for $\alpha$ in Eq.~\ref{eq:losstotal}. Setting $\alpha$ = 0 is
equivalent to the baseline model.}
\label{fig:ablation_alpha}
\end{figure*}

\medskip
\noindent

\textbf{Ablation study on the effect of transferring the shape-intensity knowledge:}
% A straightforward alternative is the trivial
% knowledge distillation that transfers the knowledge extracted by the
% model trained on the original training images. 
A straightforward alternative is to train the teacher model using the original training images.
Therefore, we train both the teacher and student U-Net model using the same setting. The
only difference with the baseline model is that we also adopt the
distillation loss when training the student model. As depicted
in Tab.~\ref{tab:input-ablation}, such a trivial alternative may
occasionally bring some intra-dataset improvement, but not as
significant as the proposed \our. 
%Besides, the generalization ability is not always improved. 
This demonstrates that the performance
improvement of \our is mainly brought by the proposed shape-intensity knowledge
transferring via knowledge distillation, not the trivial knowledge
distillation. We also conduct experiments by training the teacher
network on the annotated label maps. As depicted in
Tab.~\ref{tab:input-ablation}, the variant of \our by transferring only
the shape knowledge directly extracted from the label map is also
beneficial for improving the segmentation performance and
generalization ability. Yet, since the intensity of the label map is quite different from the original image in distribution, this variant of \our is not as
effective as \our that distills the shape-intensity knowledge from the teacher model
trained on class-wise averaged training images.

% \begin{table}[tp]
% 	\centering
% 	%   \footnotesize
% 	\caption{Quantitative results of the proposed \our on the cardiac dataset with different input for the teacher network.
% 		%$\uparrow$ (\resp $\downarrow$) indicates the higher (\resp lower) the better.
% 	}
% 	\label{tab:ACDC input ablation}
% 	% \vspace{\fixedvskip}
% 	\setlength{\tabcolsep}{2mm}{
% 		\begin{tabular}{ccc|cc}
% 			\toprule
% 			\multirow{2}{*}{Input} &
% 			\multicolumn{2}{c|}{ACDC} &
% 			\multicolumn{2}{c}{M\&Ms} \\
% 			\cmidrule{2-5}
% 		     & DICE$\uparrow$ &HD$\downarrow$  & DICE$\uparrow$ &HD$\downarrow$ \\
% 			\midrule
% 			 U-Net~\cite{ronneberger2015u} &83.06 &13.21 &63.22 &21.37\\
% 			Image &83.43 &19.05 &59.58 &23.20\\
% 			 Class-wise average &\textbf{\best{86.77}} &\textbf{\best{10.48}} &\textbf{\best{70.26}} &\textbf{\best{20.43}}\\
% 			\bottomrule
% 		\end{tabular}

% 	}
% 	%\vspace{2mm}
% \end{table}

\medskip
\noindent
\textbf{Ablation study on the loss weight $\alpha$:} The only
hyper-parameter for the proposed \our is the loss weight $\alpha$
involved in Eq.~\eqref{eq:losstotal}. We conduct ablation study on
this hyper-parameter by setting $\alpha$ to 0.5, 1.0, 2.0, 3.0, 4.0,
and 5.0, respectively. The corresponding results are depicted in
Fig.~\ref{fig:ablation_alpha}. Using different $\alpha$
slightly affects the intra-dataset performance.
% ranging from 0.890 to
% 0.894 Dice score (\resp 1.15 to 1.59 mm HD) for cardiac segmentation,
% and 0.961 to 0.963 Dice score (\resp, 0.927 to 0.929 IoU) for optic
% nerve head segmentation. 
The generalization ability changes more
significantly. \our with different settings for $\alpha$ generally
performs much better than the baseline model (equivalent to set
$\alpha = 0$), further proving the effectiveness of \our. Setting
$\alpha = 2$ gives the best Dice score on cardiac segmentation and
performs relatively well on optic nerve head segmentation. Therefore,
we set $\alpha$ to 2 for the proposed \our in all experiments.

\section{Conclusion}
\label{sec:conclusion}
In this paper, we propose a novel joint shape-intensity knowledge distillation method
for deep-learning-based medical image segmentation. 
% Differently from most methods that
% incorporate shape information via post-processing or feature
% consistency of a reconstruction network applied on the ground-truth
% segmentation and prediction, 
We leverage knowledge distillation to
transfer the shape-intensity information extracted by the teacher network trained on
class-wise averaged images. In this way, the student network with
the same network architecture as the teacher model effectively learns
shape-intensity information for medical image segmentation. 
% To the best of our
% knowledge, this is for the first time that knowledge distillation is
% applied to incorporate shape information for medical image
% segmentation. 
Extensive experiments on five medical image segmentation
tasks of different modalities demonstrate that the proposed \our
achieves consistent/significant improvements over the baseline methods
and some state-of-the-art methods. The proposed \our can be applied to
most popular segmentation network and bring performance improvement
without any additional computation effort during inference. In the
future, we would like to explore other feature layers for distillation and other distillation loss functions, and apply the proposed \our to 3D medical image segmentation task.

\section*{Acknowledgement}
This work was supported in part by the National Key Research and Development Program of China (2023YFC2705700), NSFC 62222112, 62225113, and 62176186, the Innovative Research Group Project of Hubei Province under Grants (2024AFA017), and CAAI Huawei MindSpore Open Fund.

\bibliographystyle{fcs}
\bibliography{reference}

% \vspace{-2cm}
\begin{biography}{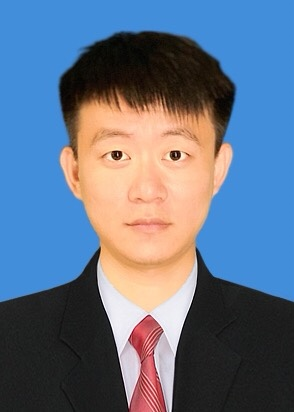}
{Wenhui Dong} received the M.S. degree from the School of Software Engineering,
Wuhan University, Wuhan, China, in 2020. He is currently pursuing the
Ph.D. degree in the School of Computer Science, Wuhan University,
Wuhan, China. His main research interests include image segmentation,
medical image analysis, and video object segmentation.
\end{biography}

\vspace{30pt}
\begin{biography}{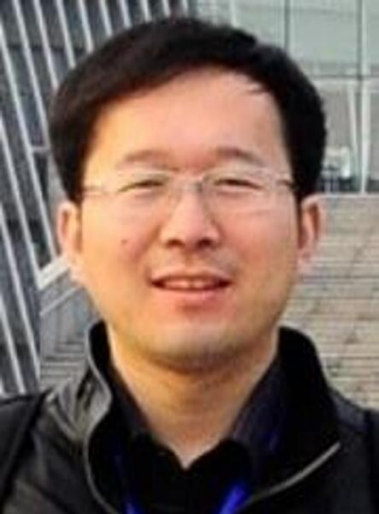}
{Bo Du}
received the Ph.D. degree in photogrammetry and remote sensing from
the State Key Lab of Information Engineering in Surveying, Mapping and
Remote Sensing, Wuhan University, in 2010. He is currently a professor
and the dean of School of Computer Science. He is also the director of
the National Engineering Research Center for Multimedia Software,
Wuhan University. His major research interests include machine
leanring, computer vision, and image processing.  He has more than 80
journal papers published in IEEE TPAMI/TIP/TCYB/TGRS, IJCV.  He serves
as associate editor of Neural Networks, Pattern Recognition, and
Neurocomputing. He won the Highly Cited researcher
(2019/2020/2021/2022) by the Web of Science Group. He also won IEEE
Geoscience and Remote Sensing Society 2020 Transactions Prize Paper
Award, and IJCAI Distinguished Paper Prize.  He regularly serves as
senior PC member of IJCAI and AAAI.
\end{biography}

\vspace{-2.5cm}
\begin{biography}{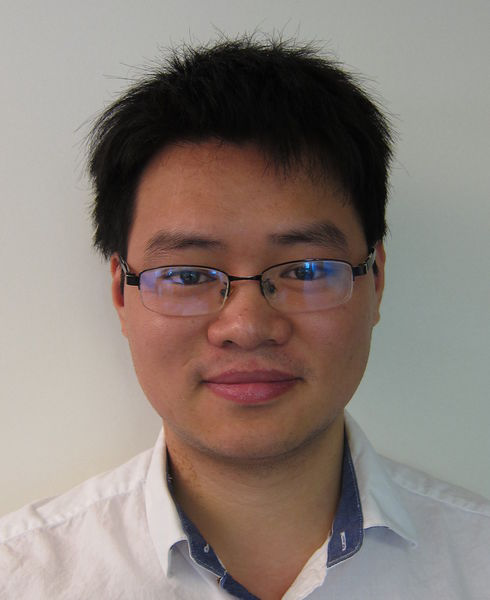}
{Yongchao Xu} received in 2010 the master degree in electronics and signal processing at Universit\'{e} Paris Sud, and the Ph.D. degree in image processing at Universit\'{e} Paris Est in 2013. He is currently a Professor with the School of Computer Science, Wuhan University. His research interests include image segmentation, medical image analysis, cross-domain generalization for deep learning.  He has published more than 40 scientific papers, such as IEEE TPAMI, IJCV, IEEE TIP, CVPR, ICCV, ECCV, MICCAI. He serves as associate editor of Pattern Recognition, Image and Vision Computing, and young associate editor of Frontiers of Computer Science. 
\end{biography}

\end{document}